%% file: Langacker-ICHEP2018.tex
\documentclass{PoS}

\usepackage{latexsym}
\usepackage{amssymb,amsmath,bm}
\usepackage{graphicx}

\newcommand{\x}{\ensuremath{\times}}                                                
\newcommand{\sto}{\ensuremath{SU(2) \x U(1)}}        
\newcommand{\Dsla}{\ensuremath{\!\not\!\!D}}
\newcommand{\lag}{\ensuremath{\mathcal{L}}}
\newcommand{\ra}{\ensuremath{\rightarrow}}
\newcommand{\snu}{\ensuremath{\stackrel{\scriptscriptstyle (-)}{\nu}}\!}

\title{Conference Summary and Outlook: Particle Physics Past, Present, and Future}
\ShortTitle{Particle Physics Past, Present, and Future}
\author{{Paul Langacker}\\ 
        Institute for Advanced Study\\ Princeton, NJ USA  08540\\
        E-mail: \email{pgl@ias.edu}}
\abstract{I briefly summarize highlights of ICHEP2018, comment on the 50$^{th}$ anniversary of the Standard Model,
and share some of my thoughts for the future.}

\FullConference{The 39th International Conference on High Energy Physics (ICHEP2018)\\
		4-11 July, 2018\\
		Seoul, Korea}

\begin{document}

\section{Introduction}
The International Conference on High Energy Physics, or ``Rochester Conference,'' has been held every one or two
years since the first one in Rochester, New York in 1950. ICHEP2018, held in Seoul, South Korea, brought
together over 1100 participants to discuss all aspects of particle physics, including
experiment, phenomenology, formal theory, astro-particle, accelerator, detector, computing, education, diversity,
applications, and ties to industry. There were some 835 parallel talks in 16 sections, 41 plenary talks, and 226 posters,
as well as satellite meetings, public lectures, award talks, and a Director's panel. I will not attempt to summarize
everything at the conference, but will focus on my own view of some of the highlights, as well as a brief
overview of the 50 year development of the Standard Model  and some thoughts for the future.

\section{Happy 50$^{th}$ birthday Standard Model!}
The ICHEP organizers asked me to begin with a brief sketch of the development of the Standard Model (SM),
which celebrated its 50$^{th}$ birthday last Fall.

\subsection{Theoretical preludes (1927-1967)}
The birth of the Standard Model is usually taken to be the publication of Weinberg's \sto\ model of leptons 
in 1967~\cite{Weinberg:1967tq}, but of course that and the subsequent additions incorporated many
ideas and experimental facts from  earlier.\footnote{For a more detailed description of the history, see, e.g.,~\cite{Pais,Weinberg:vol1,langacker:pup}.} Some of the most important ingredients were:

\begin{itemize}
\item Gauge theories. Key aspects include the quantization of the electromagnetic field in 1927~\cite{Dirac:1927dy};
the subsequent development of  quantum electrodynamics (QED) [Figure~\ref{QEDfig}], which combined quantum mechanics,
special relativity, and classical electrodynamics~\cite{QED}; renormalization theory, which evaded the infinities;
Yang-Mills theory~\cite{Yang:1954ek}, which extended gauge invariance from $U(1)$ to a non-abelian group; and
axiomatic field theory~\cite{Streater}, which clarified such fundamental field-theoretic issues as the {\em CPT} theorem. 

\begin{figure}[htbp]
\begin{center}
\includegraphics*[scale=0.8]{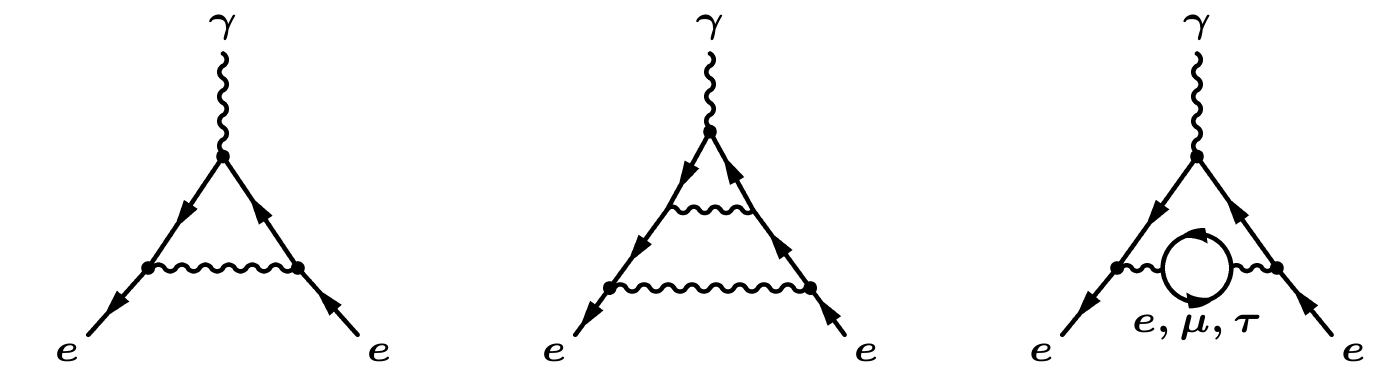}

\includegraphics*[scale=0.7]{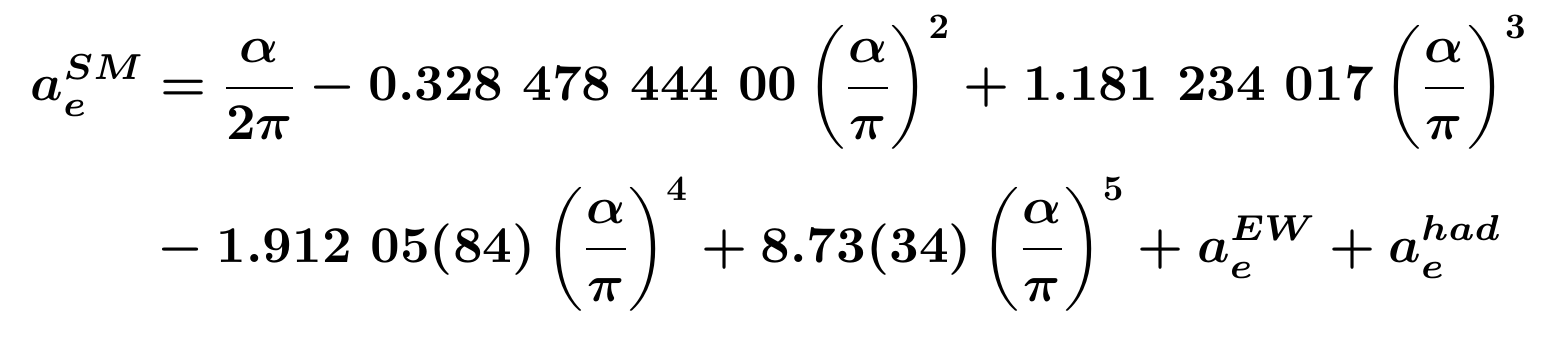}
\caption{QED has been spectacularly successful. Shown are typical  diagrams for the anomalous magnetic moment of the electron, and the SM expression to
$\mathcal{O}(\alpha^5)$.}
\label{QEDfig}
\end{center}
\end{figure}

\item The Fermi theory of the weak interactions~\cite{Fermi:1934hr}. Though not a gauge theory, the Fermi interaction (including its
subsequent modifications) provided the framework for the \sto\ theory, and is itself an excellent first approximation
to a  large variety of weak interaction processes.

\item The strong interaction approximate symmetries $SU(2)$ (isospin) and $SU(3)$ (the eightfold-way) (e.g.,~\cite{Noether:1918zz,GellMann:1962xb}) [Figure~\ref{strongsym}]; spontaneous symmetry breaking and the Goldstone theorem~\cite{Nambu:1961tp,Goldstone:1961eq}; quarks~\cite{GellMann:1964nj}; color~\cite{Greenberg:1964pe}; and scaling~\cite{Bjorken:1968dy,Breidenbach:1969kd}.

\begin{figure}[htbp]
\begin{center}
\includegraphics*[scale=0.65]{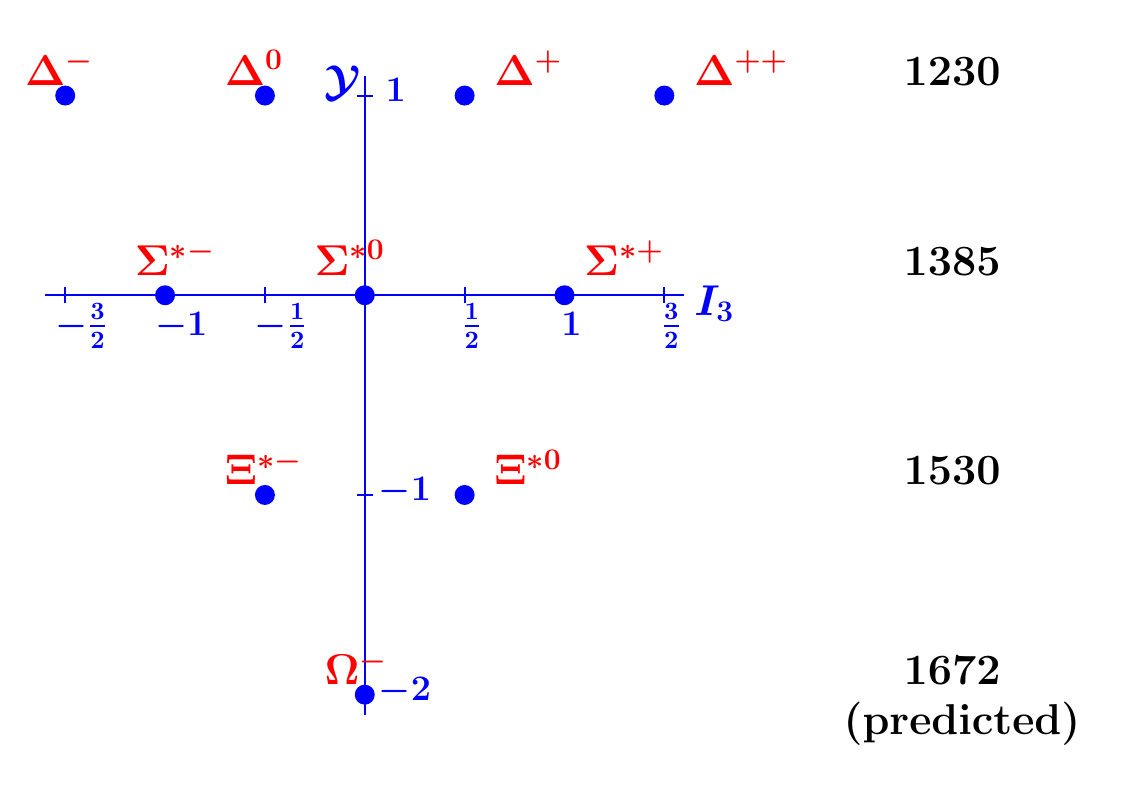}
\includegraphics*[scale=0.65]{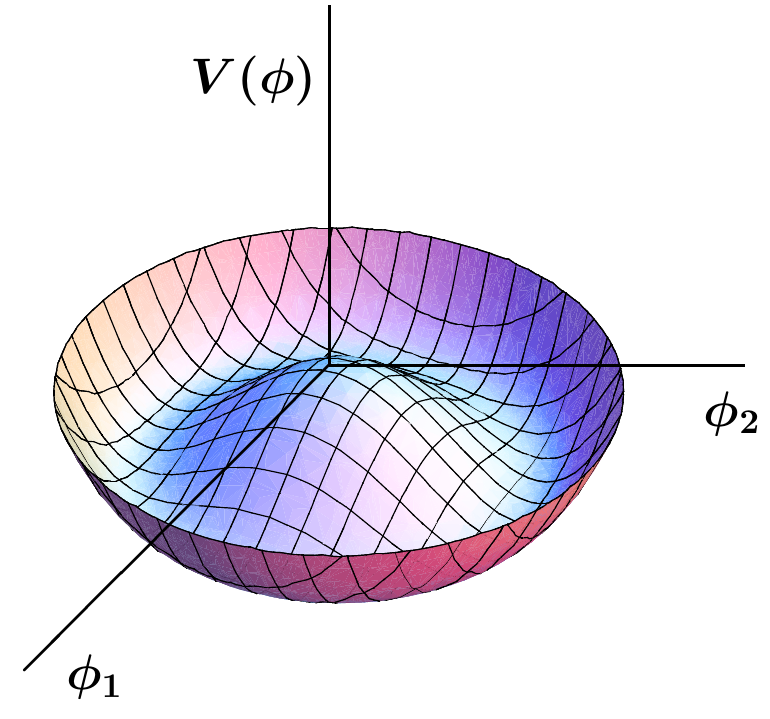}
\caption{Left: the spin-3/2 decuplet, for which the $\Omega^-$ existence and mass were successfully predicted by $SU(3)$, and which was also the original motivation for color. Right: the potential for a spontaneously broken $U(1)$ symmetry. The rolling mode around the minimum corresponds to a massless 
Goldstone boson for a global symmetry. It is ``eaten'' to become the longitudinal mode of a massive vector boson in a gauge theory. Radial excitations correspond to a massive Higgs-like boson.}
\label{strongsym}
\end{center}
\end{figure}

\item The Higgs et al. mechanism~\cite{Higgs:1964ia,Englert:1964et}, which simultaneously evaded the existence of unobserved massless Goldstone bosons in a spontaneously broken  gauge theory and allowed  nonzero mass for vector bosons.

\end{itemize}

\subsection{\boldmath The $SU(2) \times U(1)$ model of leptons}
The  Weinberg-Salam electroweak model~\cite{Weinberg:1967tq,Salam:1968rm} combined the \sto\ structure, which had been proposed previously~\cite{Glashow:1961tr}, with the Higgs mechanism, which  gave a satisfactory method for generating gauge boson masses. The Lagrangian density of the model (or any chiral gauge theory) is schematically
\[ 
 \lag = -\frac{1}{4} F_{\mu\nu} F^{\mu\nu} + \bar\psi i \Dsla \psi
+ \bigl( D^\mu \phi \bigr)^\dag \bigl( D_\mu \phi \bigr) 
-V(\phi)  - \bigl( \bar\psi_L \Gamma \psi_R \, \phi + \text{ h.c.} \bigr).
\]

Notable developments include:
\begin{itemize}
\item The original Higgs et al. papers envisioned applications (if any) to the strong interactions, whereas  in~\cite{Weinberg:1967tq} it was realized that the relevant application was  to the weak  interactions.

\item Renormalizability, originally conjectured, was established in 1971~\cite{tHooft:1971qjg}.

\item  The original model focussed on leptons because quarks were not at the time well-established. Also, a simple extension involving just three quarks ($u,\ d,\ s$) would have led to unacceptable flavor-changing neutral current (FCNC) effects. However, the gradual establishment of the quark idea and the discovery of charm ($c$) in 1974 allowed a satisfactory extension, with
flavor-changing effects cured by the GIM mechanism~\cite{Glashow:1970gm}. The subsequent discovery of the $\tau$ and $b$,
and later the $t$ quark and $\nu_\tau$, implied a third family, allowing a simple origin for $CP$ violation~\cite{Kobayashi:1973fv}.

\item The discovery of the predicted  weak neutral current (WNC) in 1973, as well as subsequent generations of
WNC experiments, and of the $W^\pm$ and $Z$ gauge bosons in 1983.

\item Precision $Z$-pole (LEP, SLC)  and collider (LEP 2, Tevatron, LHC) experiments (1989--).

\item Flavor physics, such as $K$ and $B$ decays and mixing, $CP$ violation, searches for FCNC, and the unitarity triangle.

\item The combination of these experimental probes established the basic structure of the \sto\ model, constrained small deviations from many types of new physics, confirmed the program of renormalization, correctly predicted the approximate values of the $t$ and (less accuratedly) Higgs masses, and indicated an approximate unification of the gauge couplings (including the QCD coupling) at a very high energy scale. (The gauge unification is more successful in the supersymmetric extension.) [Figure~\ref{smtests}]

\begin{figure}[htbp]
\begin{center}
\begin{minipage}{6.0cm} 
\hspace{-2.5cm}
\includegraphics*[scale=0.35]{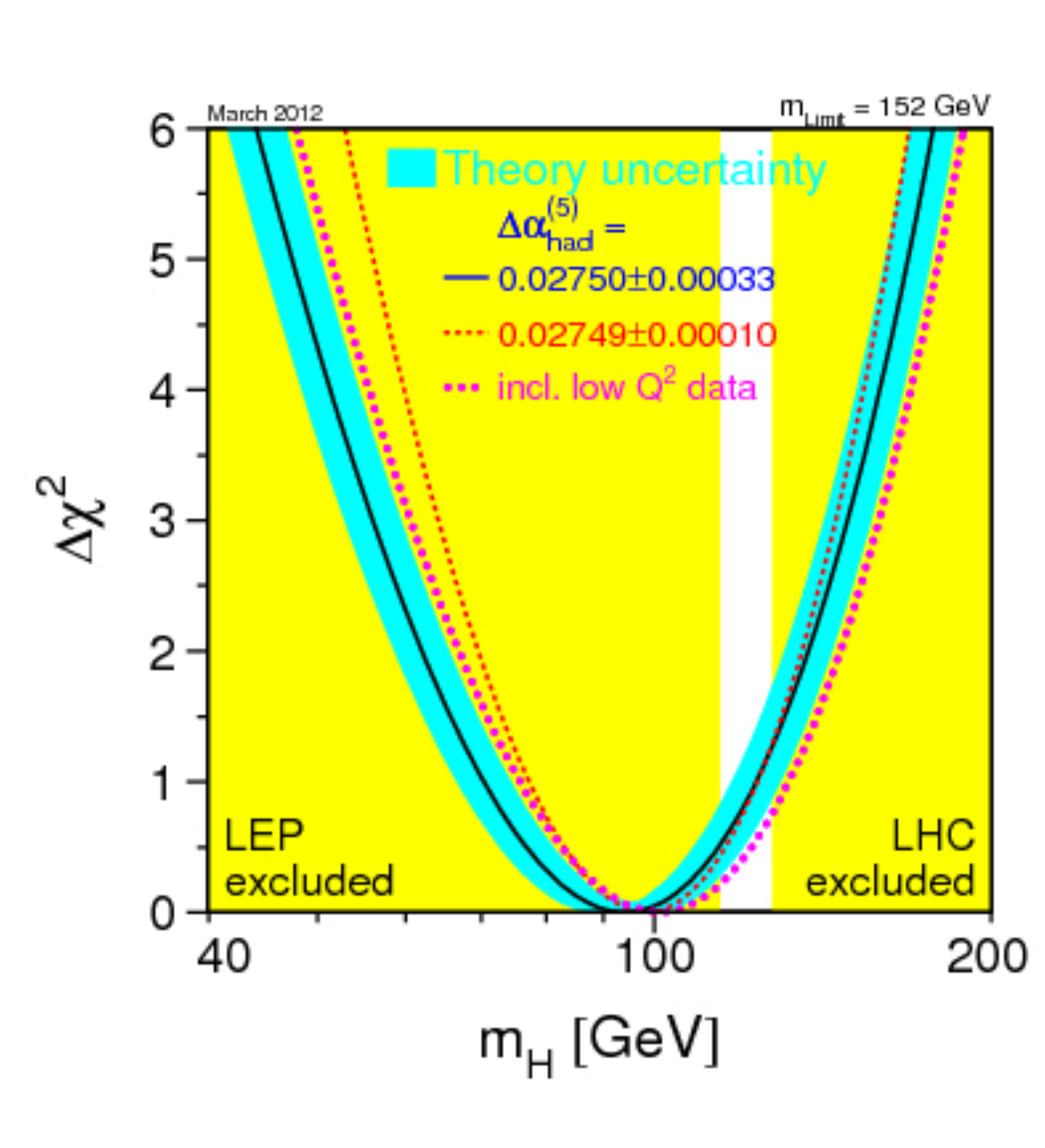}
\vspace{-2cm}
\end{minipage}
\hspace*{-1.8cm}
\begin{minipage}{5.5cm} 
\includegraphics*[scale=0.4]{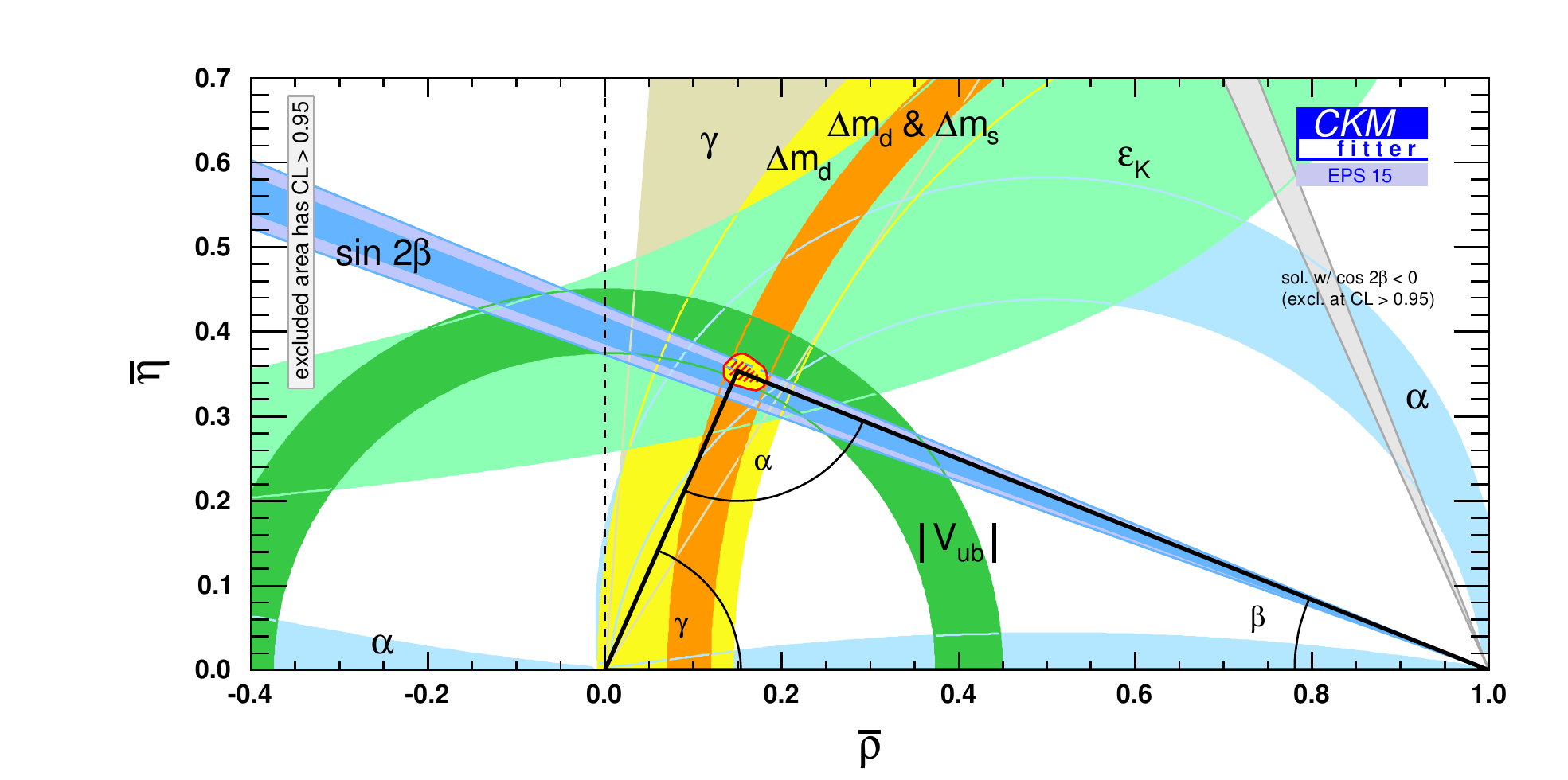}

\hspace{.5cm}\includegraphics*[scale=0.45]{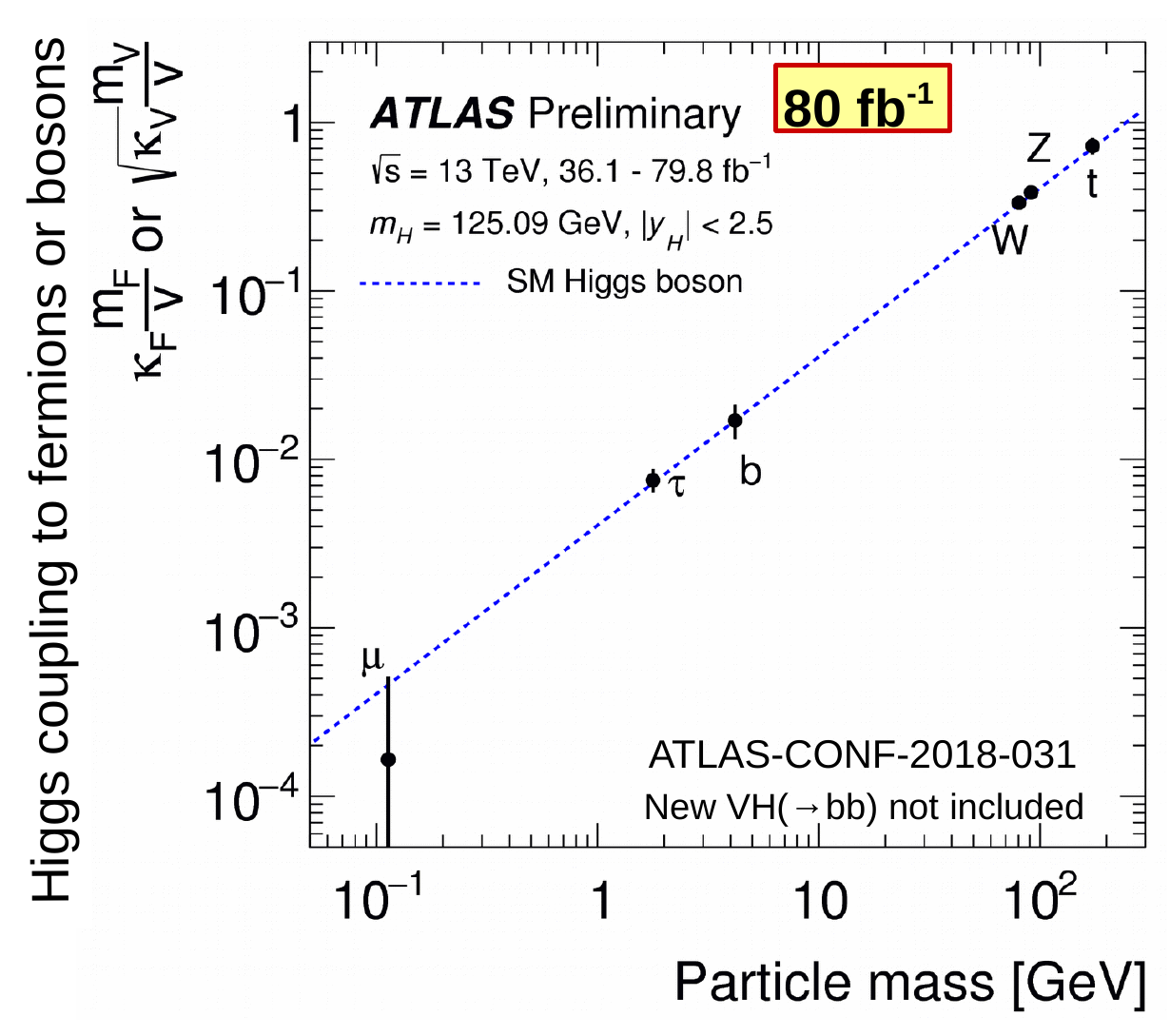}
\end{minipage}
\caption{Left: The ``blue-band'' plot predicting the Higgs mass from indirect precision data, courtesy
of the LEP Electroweak Working Group~\cite{Schael:2013ita} [{\tt http://lepewwg.web.cern.ch/LEPEWWG/}]. Top right, the unitarity triangle, courtesy of the CKMfitter group~\cite{Charles:2015gya} [{\tt http://ckmfitter.in2p3.fr/}].
Bottom right: observed and predicted Higgs couplings~\cite{Carli}.}
\label{smtests}
\end{center}
\end{figure}

\item The observation of neutrino mass and mixing effects,\footnote{The original Standard Model assumed massless neutrinos, and many authors consider neutrino mass to be ``Beyond the Standard Model'' (BSM). I choose instead to redefine the Standard Model to include neutrino mass in the same way that it was redefined to included the third family of fermions.} established in atmospheric neutrino oscillations
in 1998~\cite{Fukuda:1998mi}  (but with strong indications earlier from Solar neutrinos). It is still not certain whether the masses are Dirac or Majorana.

\item The observation of the Brout-Englert-Higgs (BEH) or ``Higgs'' boson by the ATLAS and CMS collaborations in 2012.

\end{itemize}

\subsection{Quantum Chromodynamics (QCD)}
The strong interaction part of the  Standard Model is QCD [Figure~\ref{QCDfig}], which was developed in the early 1970s
(e.g.,~\cite{Fritzsch:1973pi}). QCD is a gauge theory based on $SU(3)$. It is essentially the unique renormalizable field theory
consistent with the experimental information available at the time.
Principal ingredients and developments included:
\begin{itemize}
\item Deep inelastic $ep$ scattering~\cite{Breidenbach:1969kd} and $e^+e^-$ annihilation experiments at SLAC, which established spin-1/2 partons (quarks) and color.

\item The notions of asymptotic freedom~\cite{Gross:1973id,Politzer:1973fx}, infrared slavery, and color confinement. 

\item The observation of the gluon at DESY in 1979.

\item Sophisticated theoretical work involving perturbative QCD,  jet physics, hadronization,  the development of generators, and amplitude methods.

\item First principle lattice QCD calculations of the hadron spectrum, confinement, weak matrix elements, etc.

\begin{figure}[htbp]
\begin{center}
\includegraphics*[scale=0.6]{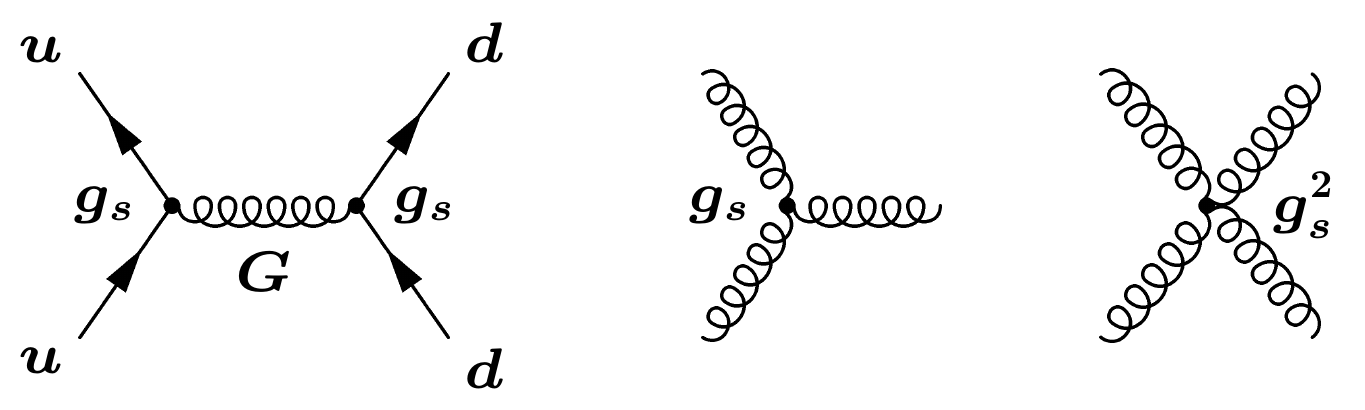}

\includegraphics*[scale=0.5]{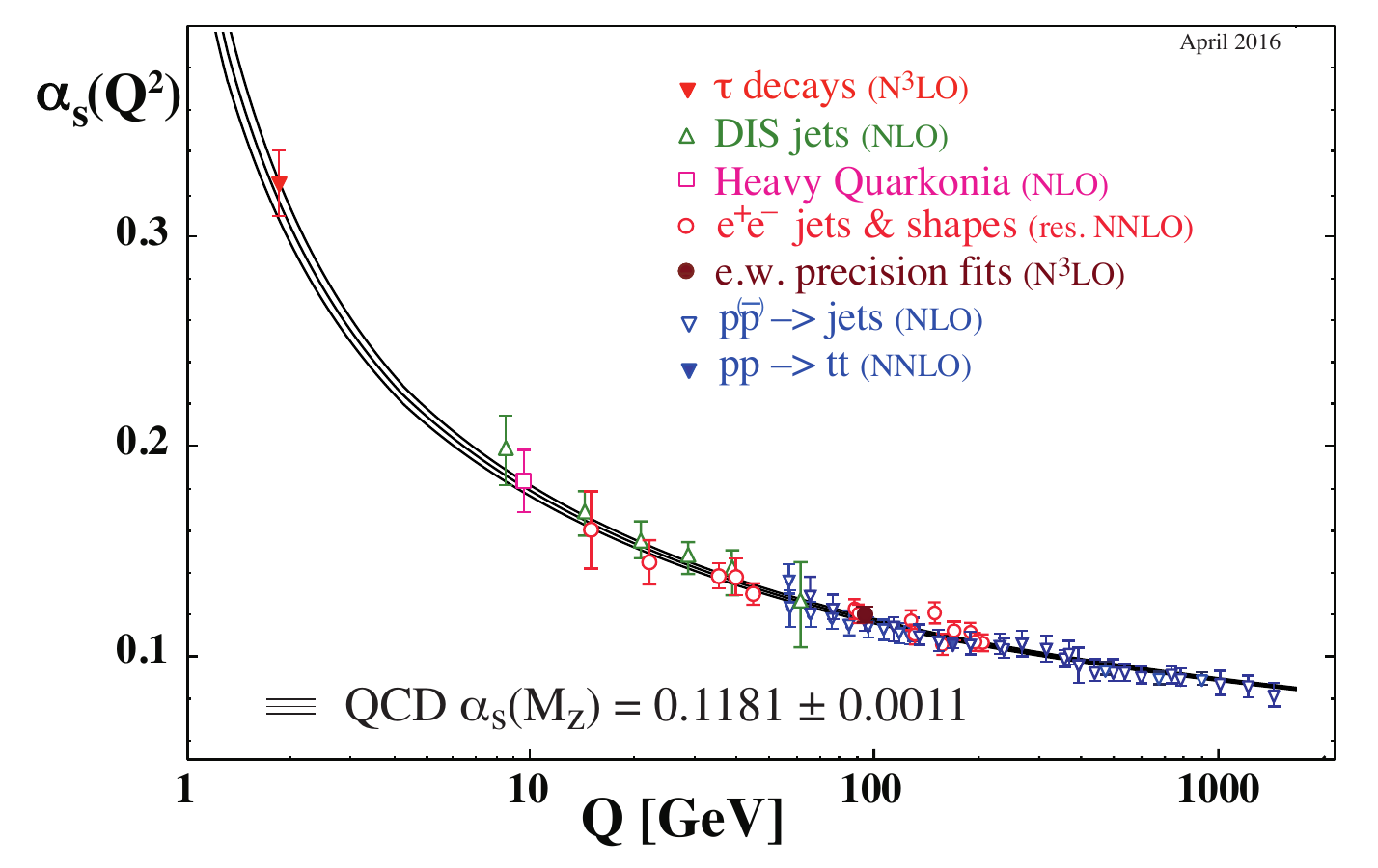}
\caption{Top: gluon exchange between quarks, and the gluon self-interactions. Bottom: the running QCD coupling, courtesty of the Particle Data Group [{\tt  http://pdg.lbl.gov}].}
\label{QCDfig}
\end{center}
\end{figure}

\end{itemize}

\subsection{Standard Model successes, problems, missing ingredients, and questions}\label{smsummary}
The Standard Model has been spectacularly successful, far more so than (hardly anyone) anticipated 50 years ago.
It is mathematically consistent and describes ordinary matter and interactions to an excellent approximation down to a distance scale of some $10^{-16}$ cm. It makes many predictions and has successfully passed many experimental tests, often to high precision.
Furthermore, its gauge and matter content leads to several  accidental symmetries at the perturbative/renormalizable level, such 
as baryon and lepton number conservation, and the absence of tree-level flavor changing neutral currents and electric dipole moments (EDMs).\footnote{With the appropriate caveats for some extensions involving Majorana neutrino masses and the strong $CP$ parameter.}

Nevertheless, there are a number of problems, missing ingredients, and questions, including:
\begin{itemize}
\item There are 27 (29) free parameters for Dirac (Majorana) neutrinos.
\item The gauge interactions are very complicated, involving the direct product of three groups with different properties.
\item There is no complete explanation for charge quantization.\footnote{Anomaly cancellation must be supplemented with additional assumptions, such as family universality.}
\item There is no fundamental explanation for the fermion families, masses, and mixings (including the type and magnitudes of the neutrino masses). These account for most of the free parameters.
\item There are severe fine tunings, including the Higgs mass and associated electroweak scale, vacuum energy (cosmological constant), and $\theta_{\text{QCD}}$.
\item The Standard Model does not include  quantum gravity.
\item The initial conditions on the Big Bang.
\item The origin of the baryon asymmetry.
\item The nature of the dark matter and energy.
\item Such conventional paradigms as uniqueness, naturalness, and minimality should possibly be reconsidered.
\item 
Why does Nature appear to be ``just right''?
\item There are possible experimental deviations from the SM, such as $g_\mu-2$, the $B$ decay anomalies, and possible sterile neutrinos.	
\end{itemize}

\section{ICHEP2018}
I now turn to a selection of highlights from the conference and of developments in the two years since ICHEP2016.

\subsection{Collider physics}
\begin{figure}[htbp]
\begin{center}
\includegraphics*[scale=0.35]{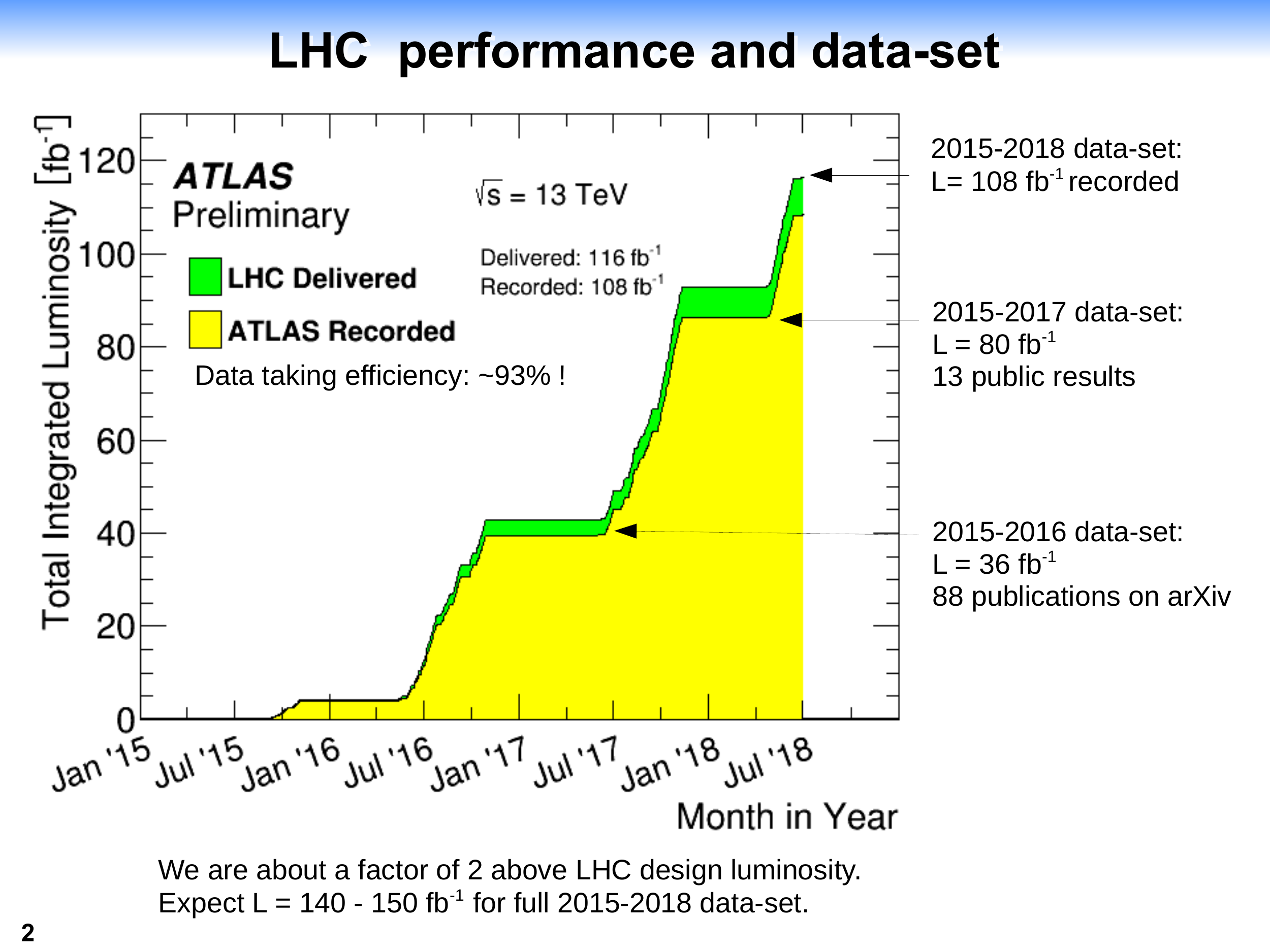}

\includegraphics*[scale=0.25]{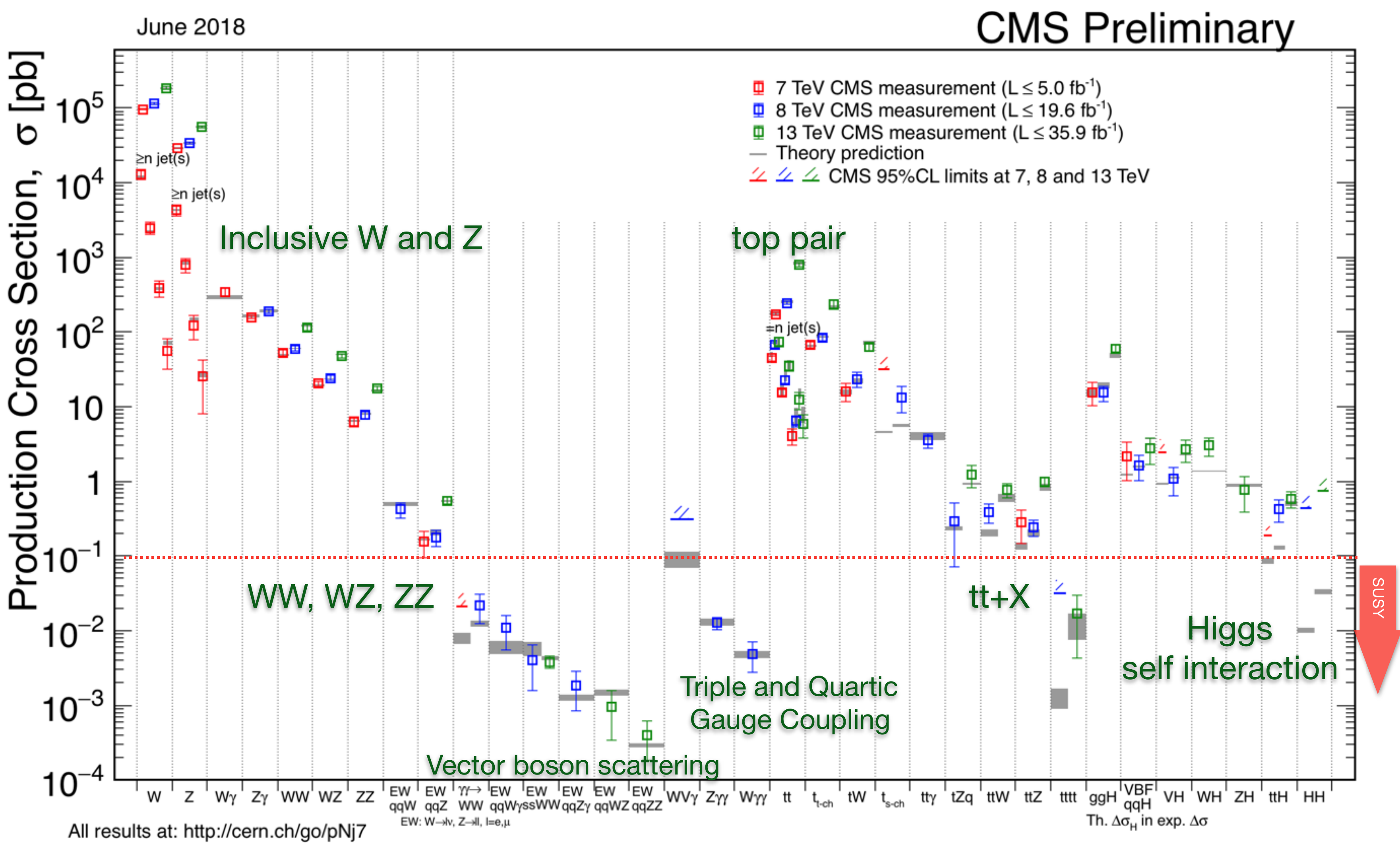}
\caption{Top: LHC luminosity~\cite{Carli}. Bottom: Observed cross sections compared with SM expectations~\cite{Rahatlou}.}
\label{collider}
\end{center}
\end{figure}

Although no major discoveries were reported, the technical accomplishments in collider physics were impressive [Figure~\ref{collider}]:
\begin{itemize}
\item The LHC and detectors have performed superbly~\cite{Carli,Rahatlou}. The luminosity for ATLAS and CMS is around twice the design value, with average pileups around 33! Major detector upgrades have been accomplished, and numberous advances in analysis methods and machine learning have been carried out.
\item There have been major programs in perturbative QCD, with most important processes computed to NNLO. There have been corresponding advances in PDFs, jet substructure, generators, amplitude methods, and lattice calculations.
\item A wide range of QCD and electroweak processes have been measured, with impressive agreement with
Standard Model expectations.
\item Phase II of the SuperKEKB/BELLE II project has started running~\cite{Akai}, with a luminosity 40 times that of KEK.

\end{itemize}

\subsection{The Higgs}
The observed 125 GeV  boson is consistent with the (minimal) SM Higgs. However, its couplings have only been measured at the 10-25\% level and more complicated scenarios are still possible. These include two or more Higgs doublets (as in supersymmetry), 
 the existence of additional scalar singlets or triplets (as in many models of electroweak baryogenesis or low-scale neutrino seesaws),
 or composite Higgs models. 
 
\begin{figure}[htbp]
\begin{center}
\includegraphics*[scale=0.55]{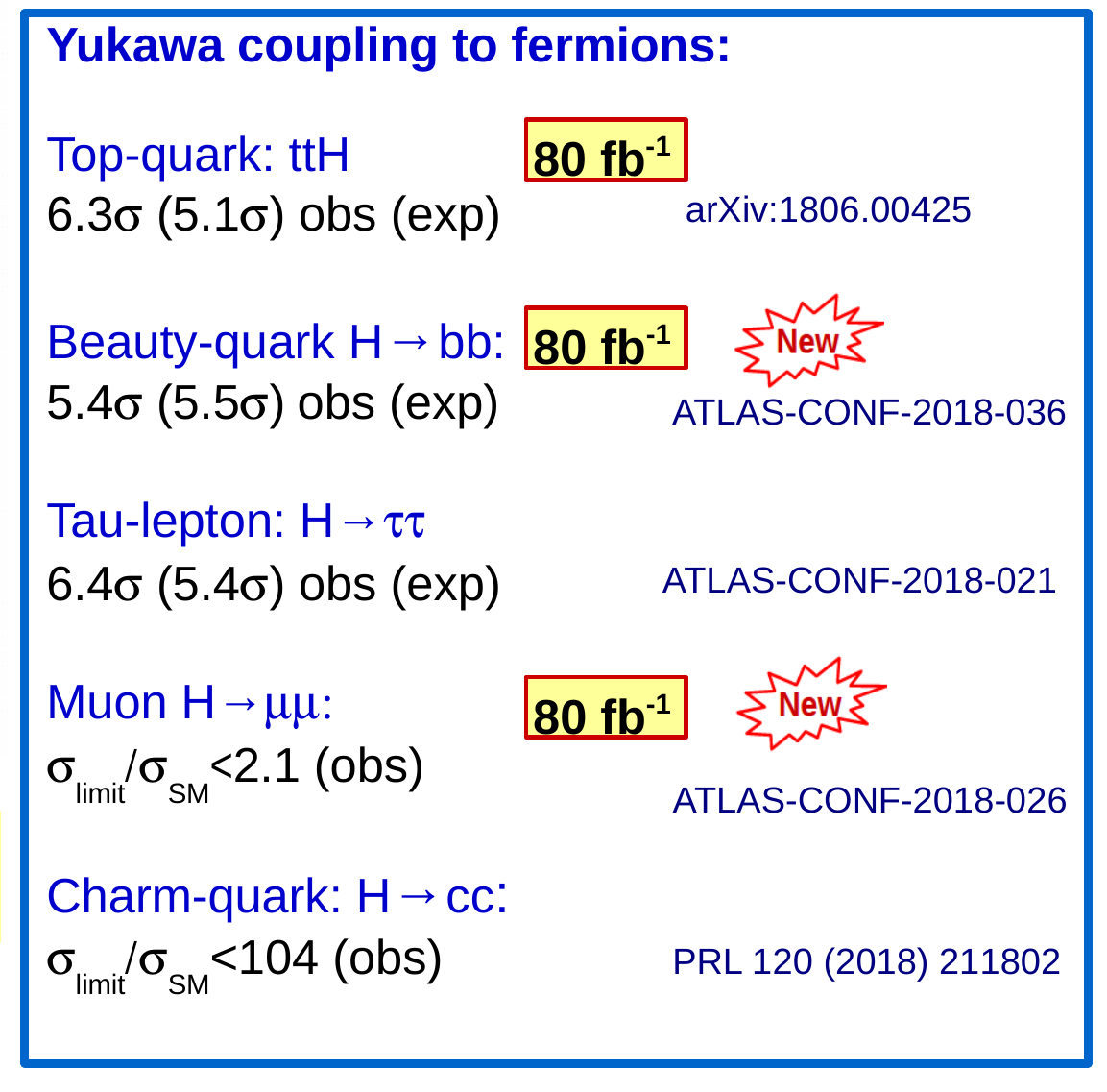}
\hspace{1cm}
\includegraphics*[scale=0.4]{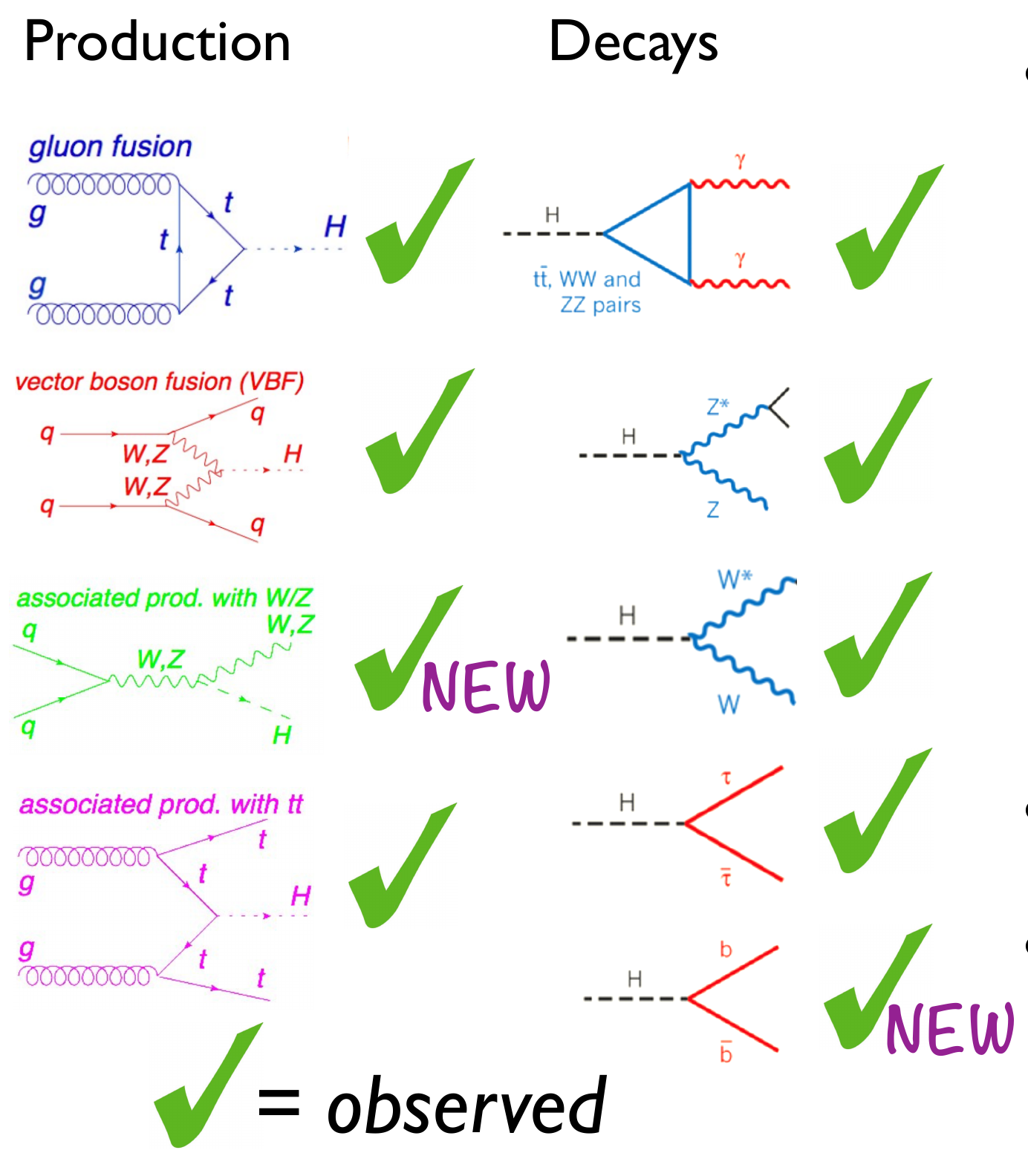}
\caption{Left: Higgs Yukawa couplings (ATLAS)~\cite{Carli}. Right: Higgs production and decay modes~\cite{Piacquadio}.}
\label{Higgscoup}
\end{center}
\end{figure}

 ATLAS and CMS have continued to refine their measurments of the BEH boson mass, couplings, and properties~\cite{Piacquadio} [Figures~\ref{smtests} and~\ref{Higgscoup}].
 In particular, the $t\bar{t}H$, $b\bar{b}H$, and $\tau^+\tau^-H$ Yukawa couplings have been observed at more than 5$\sigma$, consistent with the SM, and a non-trivial upper limit placed on  $H\rightarrow \mu^+\mu^-$  ($\sigma < 2.1 \sigma_{\text SM}$).
 All major production modes have been observed, ATLAS has set a new indirect limit of 14.4 MeV on the width (vs. 4.1 MeV in the SM), $p_T$ and other distributions have been measured, and limits have been set on additional Higgs particles.
 
 Alternatives/extensions to the minimal model often predict deviations in the couplings at the 1-10\% level.
 The HL-LHC and proposed future $e^+e^-$ and $pp$ colliders are expected to have $\mathcal{O}$(1\%) precision, and may be able to measure the induced $H^3$ vertex (which is especially sensitive to new physics)
and the total width (via $e^+e^-\rightarrow Z^\ast\rightarrow Z[H\rightarrow ZZ^\ast]$, combined with branching ratios).
  
\subsection{Heavy ions}
Heavy ion collsions~\cite{Averbeck,Hirano,Carli} explore the physics of QCD at extreme  temperatures and densities, with
implications for QCD itself as well as for conditions in the early Universe (the ``quark-gluon plasma'' or QGP) and in dense astrophysical objects.

\begin{figure}[htbp]
\begin{center}
\includegraphics*[scale=0.42]{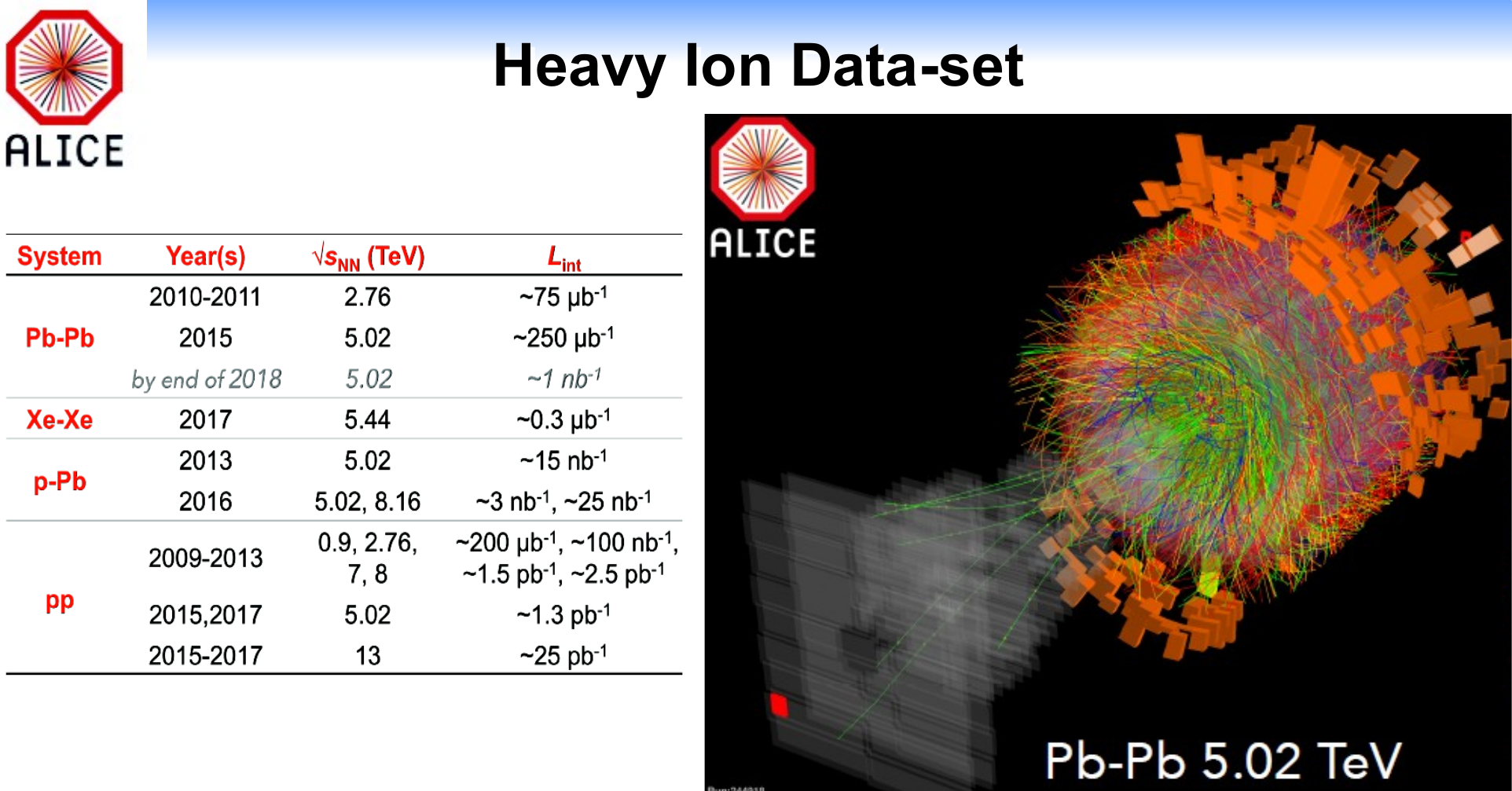}
\hspace{1cm}
\includegraphics*[scale=0.42]{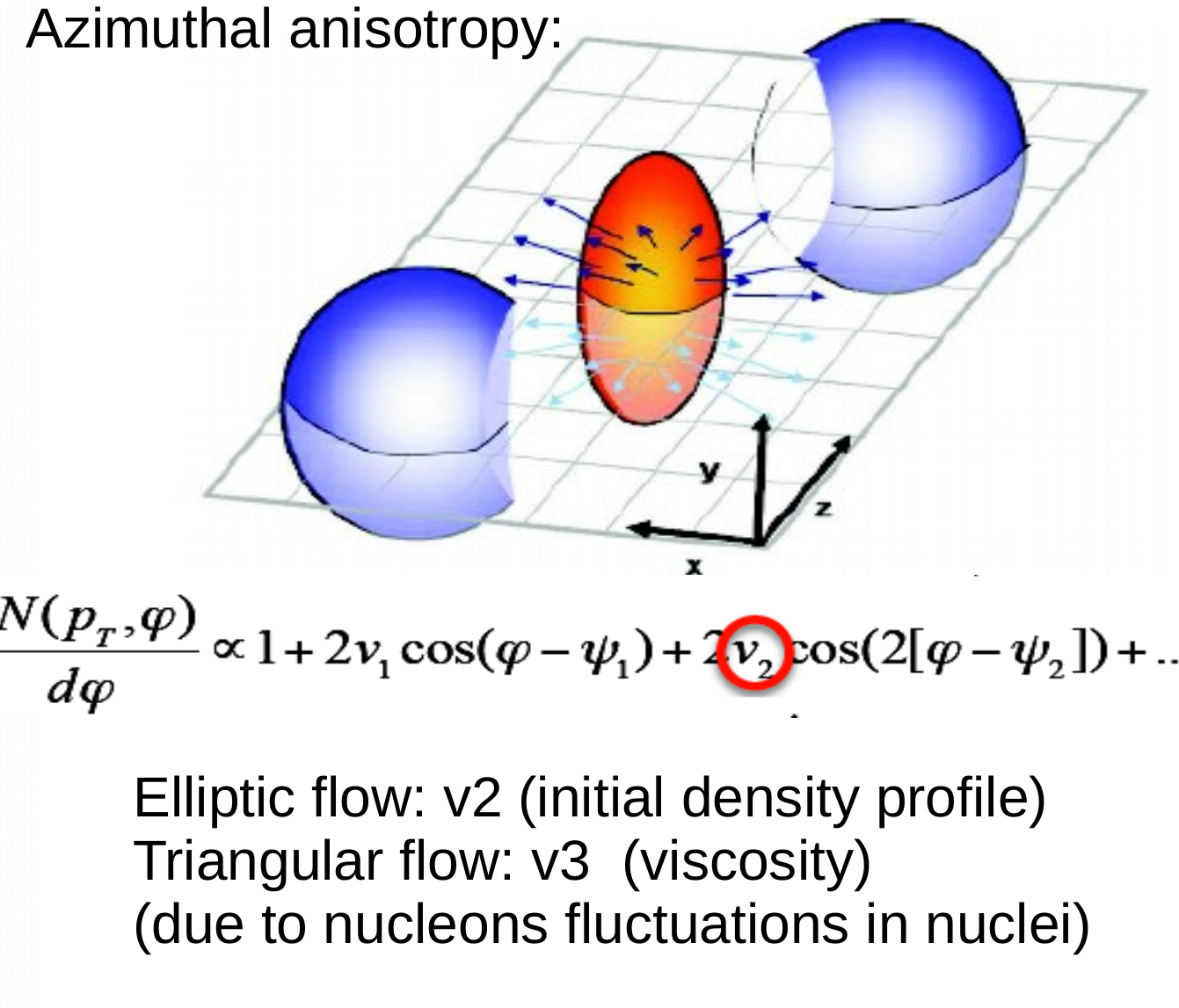}
\caption{Left: ALICE heavy ion data~\cite{Carli}. Right: Azimuthal flows~\cite{Carli}.}
\label{heavyion}
\end{center}
\end{figure}

 There has been an extensive experimental program at RHIC at Brookhaven,
and by the
 ALICE, CMS, and ATLAS experiments at the LHC [Figure~\ref{heavyion}]. The LHC running has extensively studied hot (``fireball'') QGP effects in $Pb-Pb$ collisions (and  some recent running of $Xe-Xe$), as well as using   $pp$ and $p-Pb$  data as baselines to separate ordinary hadronic and nuclear effects. Observations include the (very large) multiplicities as functions
 of the centrality of the collision, $p_T$, and particle type; jet quenching and propagation of heavy flavors; long-range correlations; and azimuthal flows. The results imply  that under existing collider conditions the final state is neither
 a collection of hadrons nor an ideal gas of quarks and gluons. Rather, there are strong collective effects more characteristic of a smoothly flowing fluid.

\subsection{Electroweak and top} 
\begin{figure}[htbp]
\begin{center}
\includegraphics*[scale=0.4]{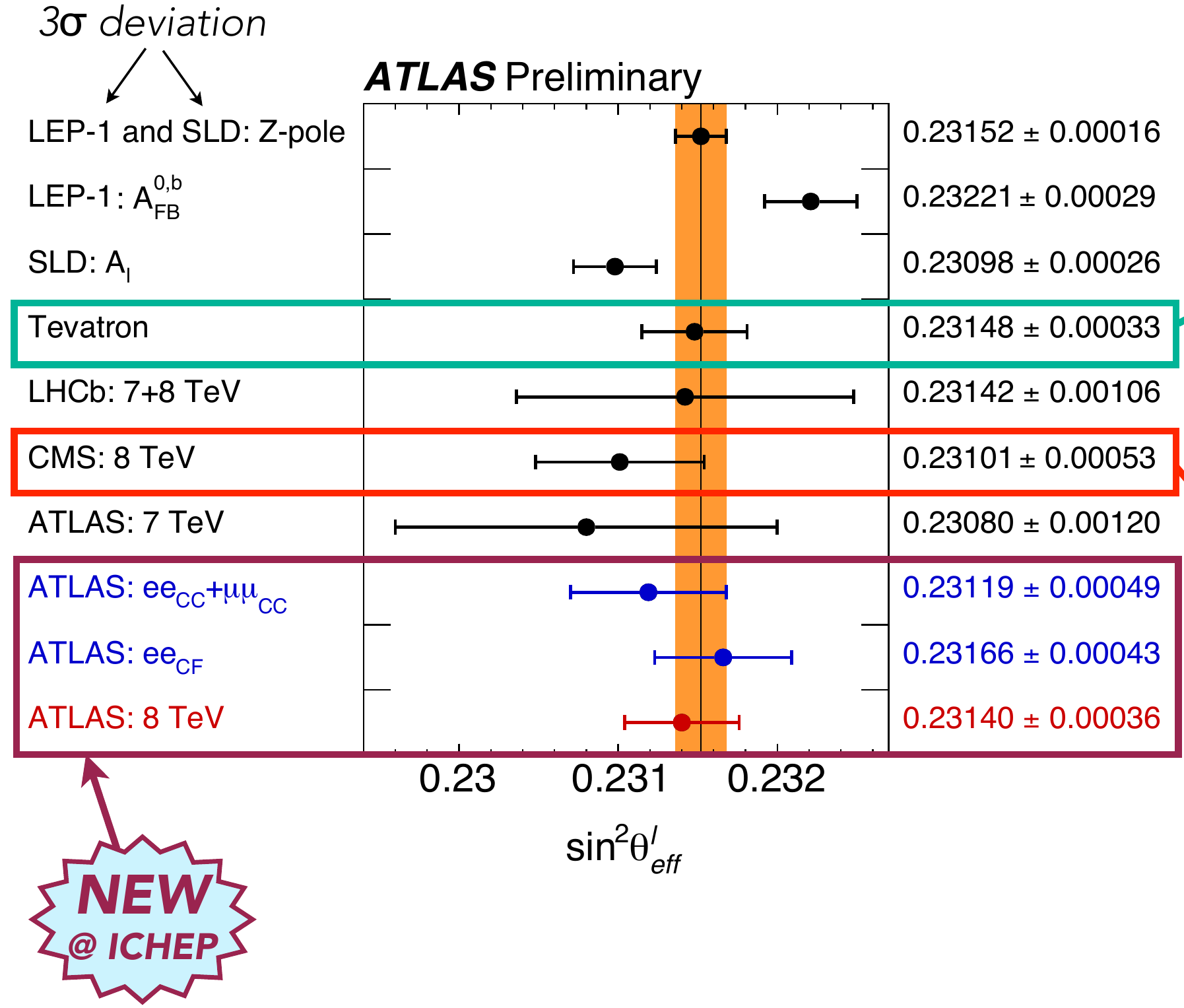}

\includegraphics*[scale=0.3]{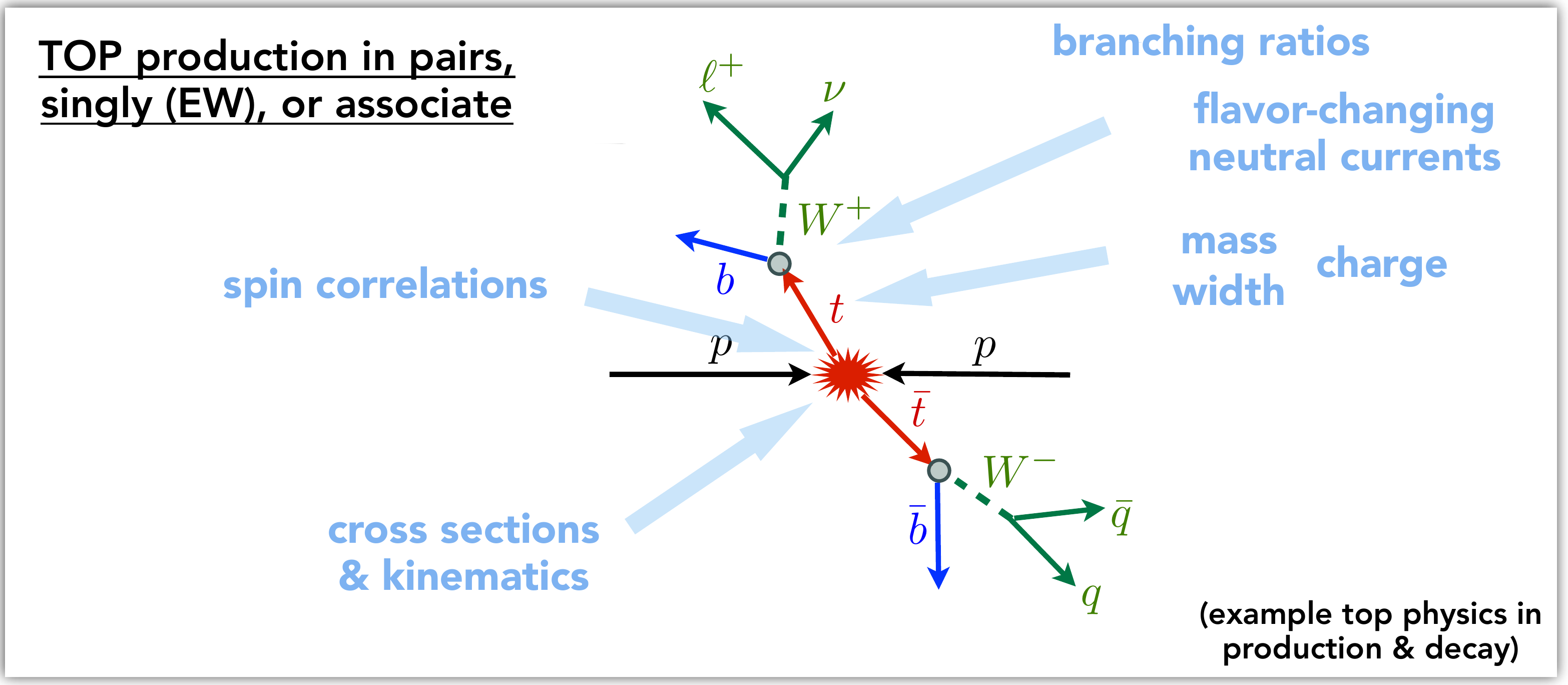}
\caption{Top: Measurements of the weak angle~\cite{Skinnari}. Bottom: Top-pair production~\cite{Skinnari}.}
\label{electroweak}
\end{center}
\end{figure}

Electroweak measurements at the LHC~\cite{Skinnari} include precise determinations of the leptonic
weak angle $\sin^2 \theta^\ell_{eff}$ from asymmetries in dilepton production (especially important
because of the discrepancy between the most precise $Z$-pole determinations) [Figure~\ref{electroweak}]; $W$, $Z$, and $\gamma$ production (as tests of QCD, $M_W$, etc.); $WW$, $WZ$, and other diboson productions (as tests of triple gauge vertices); 
and diboson scattering (as tests of unitarity, the electroweak symmetry breaking mechanism, and quartic gauge vertices).

Top quarks are produced prolifically in $t\bar{t}$ pairs (180 million so far at the LHC~\cite{Dolan}), as well as singly or in association with other particles~\cite{Skinnari,Dolan}. Measurements of  cross sections, spin correlations, etc. are important as QCD tests, for  a precise  $m_t$
(needed for precision electroweak and Higgs physics), determination of $V_{tb}$, and searches for new physics (such as in rare flavor-changing decays). High precision calculations of SM expectations, such as NNLO QCD and NLO EW, are essential to match the experimental precision~\cite{Dolan}.

\subsection{Exotic searches}
\begin{figure}[htbp]
\begin{center}
\hspace{-1.5cm}
\begin{minipage}{5.5cm} 
\includegraphics*[scale=0.40]{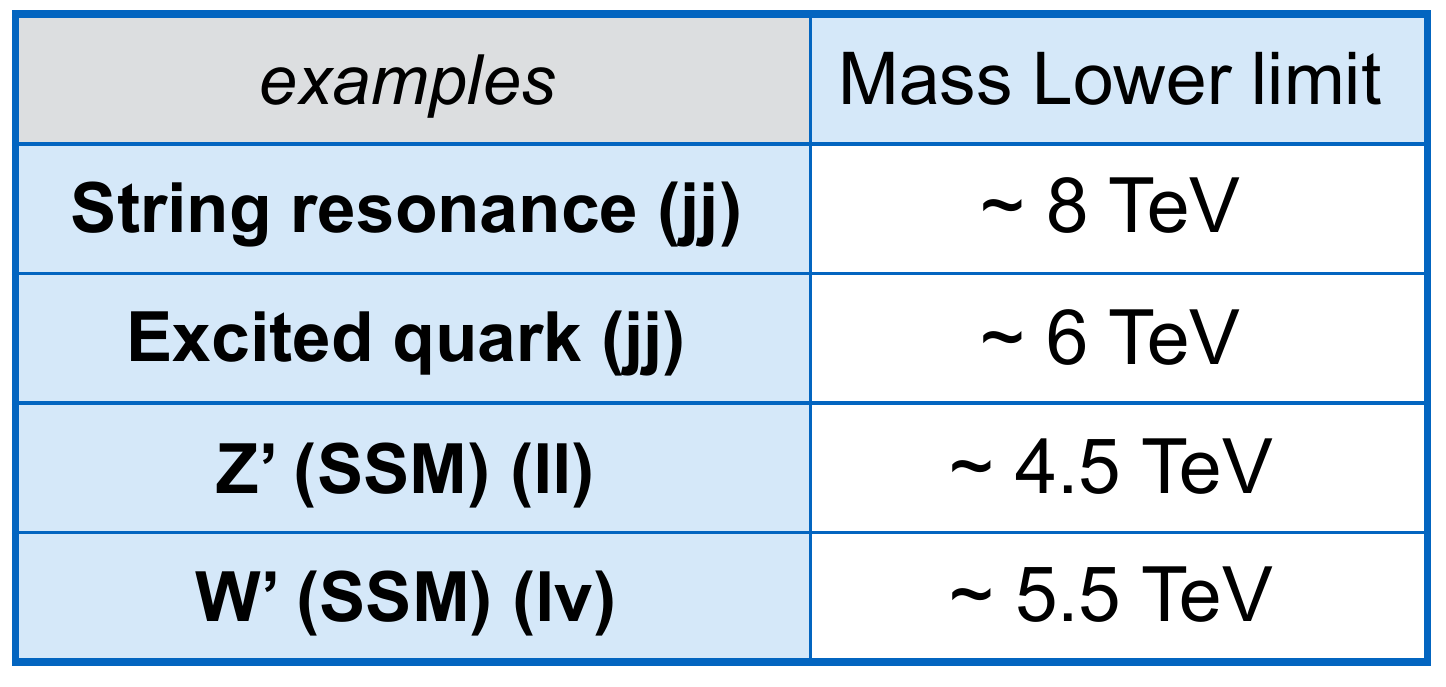}
\end{minipage}
\hspace{0.5cm}
\begin{minipage}{6.0cm} 
\includegraphics*[scale=0.45]{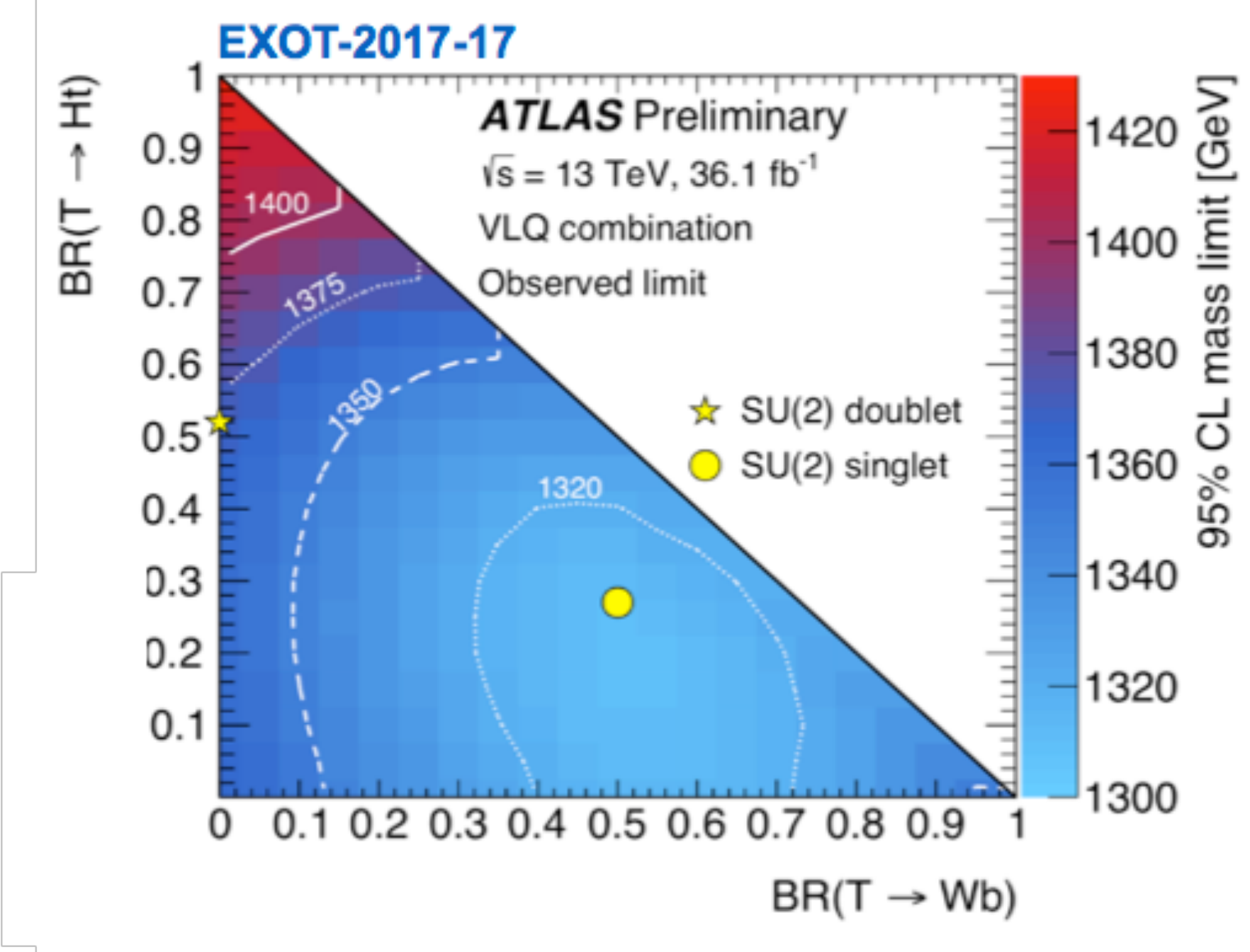}
\end{minipage}
\caption{Left: Limits on resonances~\cite{delray}. Right: $T\bar{T}$ pairs~\cite{delray}.}
\label{exotics}
\end{center}
\end{figure}

Searches for exotics~\cite{delray} include vector (non-chiral under the SM) fermion pairs, such as heavy charge-2/3 quarks $T-\bar{T}$ decaying into ordinary quarks plus $W$, $Z$, or $H$; heavy resonances, such as $Z'$, $W'$, colorons, etc.
(searched for in dilepton, dijet, diboson, or $t\bar{t} $ spectra); weakly coupled or long-lived particles; dark matter; leptoquarks (suggested in some explanations of the $B$ anomalies);
diquarks; and particles that could be portals to a dark sector, such as extra Higgs particles or light weakly-coupled vectors.
Some  of these, such as vector fermions or $Z'$ emerge in many types of new physics (e.g., as string remnants or in composite Higgs models), which is both an advantage (more likely to occur) and disadvantage (harder to diagnose if observed) [Figure~\ref{exotics}].

New or improved search techniques include boosted jets/jet substructure, associated production, and mono-particles associated with missing transverse energy.

\subsection{Supersymmetry}
Supersymmetry has long been one of the most popular extensions of the SM,
motivated by the Higgs naturalness problem, gauge coupling unification, and its prediction of a plausible WIMP
candidate for cold dark matter. Furthermore, the Coleman-Mandula theory asserts that under plausible assumptions supersymmetry
is the unique space-time  extension of the Poincar\'e algebra. Finally, supersymmetry is required in realistic string constructions (although the supersymmetry breaking scale could be large).

There are many versions of supersymmetry, even within the minimal model (MSSM).
Early experimental searches focussed on favorable scenarios, e.g., involving large missing $E_T$, $R$-parity ($R_P$) conservation, etc., and often utilized  specific supersymmetry-breaking models, or simplified models in which there was a single decay mode. Since no signals have been observed, most recent analyses\footnote{Most of the results presented at ICHEP2018 utilize $\sim36$ fb$^{-1}$/detector of 13 TeV data from 2015-2016,
with some  using 80 fb$^{-1}$ from 2015-2017.}~\cite{Strandberg} have concentrated on more challenging possibilities, e.g., with less missing $E_T$, heavy neutralinos, multiple decays modes, or longer decay chains. These are often analyzed in the framework of the much more
general phenomenological MSSM (pMSSM), or in specific scenarios such as compressed, stealth, split, light gravitino (gauge mediation), $R_P$ violation, etc. There have also been many more studies of electroweak sparticles and of the third sfamily, which have lower production cross sections. 

\begin{figure}[htbp]
\begin{center}
\includegraphics*[scale=0.6]{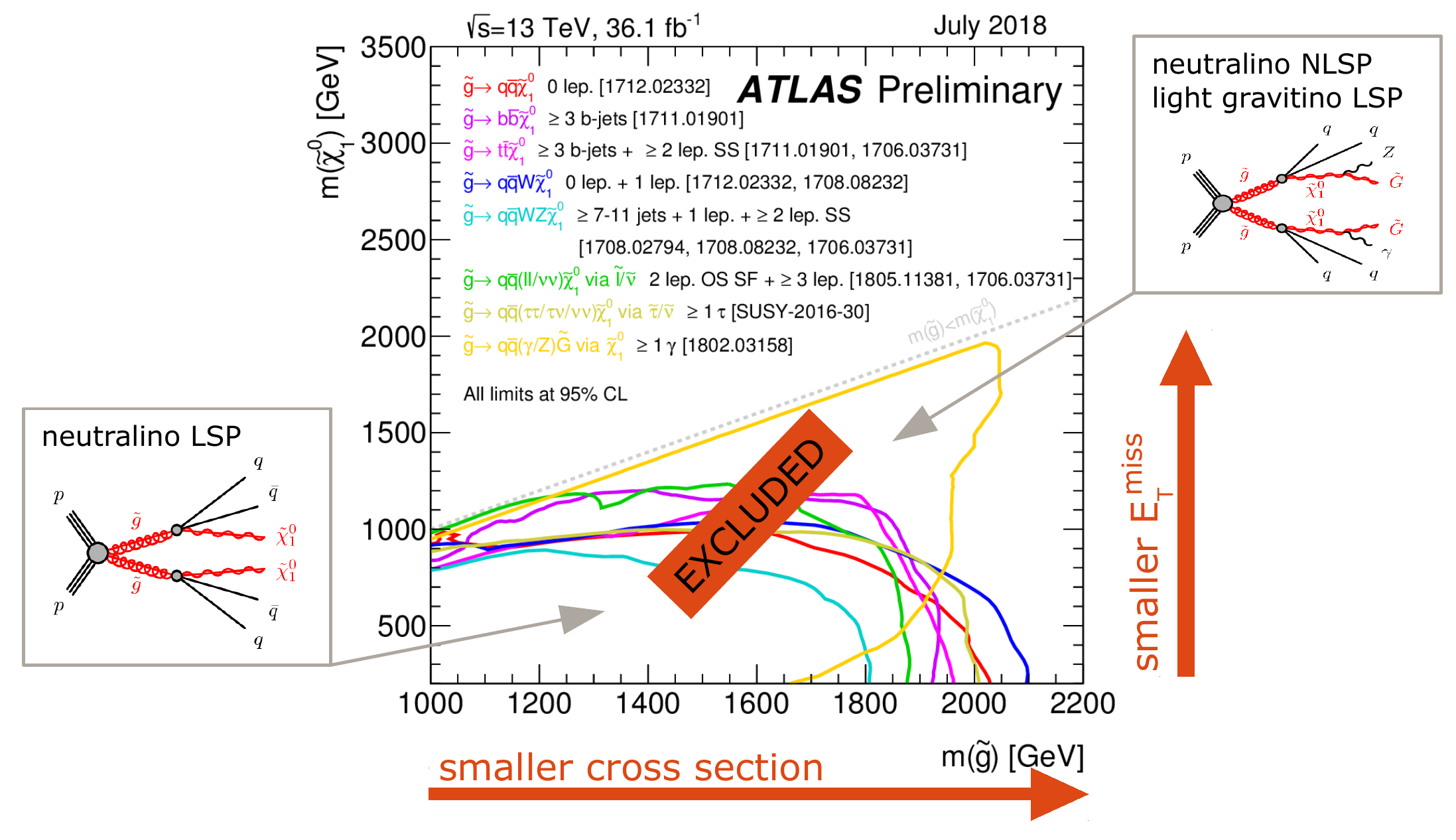}
\caption{Typical exclusions in the MSSM~\cite{Strandberg}.}
\label{susy}
\end{center}
\end{figure}

 So far, there is no sign of supersymmetry [Figure~\ref{susy}]. A  rough summary of the current exclusions under
 favourable/challenging scenarios is~\cite{Strandberg}:
 \begin{itemize}
\item gluinos up to $\mathcal{O}(2)/\mathcal{O}$(1) TeV.
\item Squarks up to $\mathcal{O}(1.5)/\mathcal{O}$(0.5) TeV.
\item Stops and sbottoms up to $\mathcal{O}(1)/\mathcal{O}$(0.7) TeV.
\item EW produced sparticles  up to $\mathcal{O}(0.5-1)/\mathcal{O}$(0.1) TeV.
\end{itemize}

 Although supersymmetry may still very well turn up in future analyses and runs, the lack of evidence for this or many other popular extensions of the SM is beginning to challenge the naturalness paradigm.
 Also,  a higher scale for the superpartner masses makes it difficult to explain the $g_\mu-2$ anomaly using supersymmetry.
 On the other hand, it somewhat relaxes the tension from the nonobservation of FCNC or EDMs and makes it easier to accomodate the observed Higgs mass (which is a bit large for much of the MSSM parameter space).
 
 There are also many plausible extensions of the MSSM, e.g., involving additional $SU(2)$-singlet Higgs fields (the NMSSM and generalizations) or the UMSSM (involving an additional $Z'$), which often relax the naturalness issue.

\subsection{Flavor physics}
\subsubsection*{$\boldmath CP$ violation and the CKM matrix}
\begin{figure}[htbp]
\begin{center}
\includegraphics*[scale=0.45]{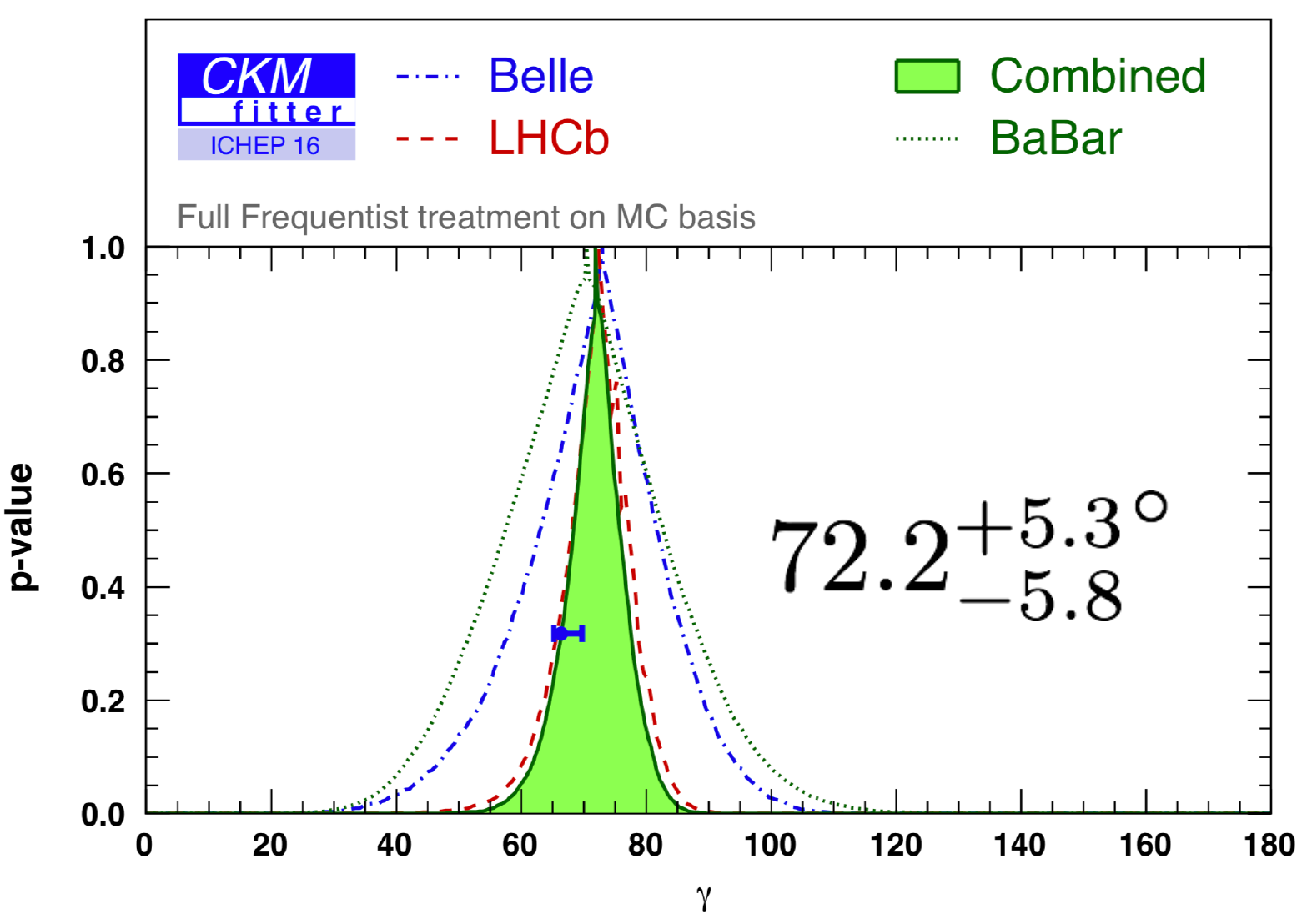}
\includegraphics*[scale=0.37]{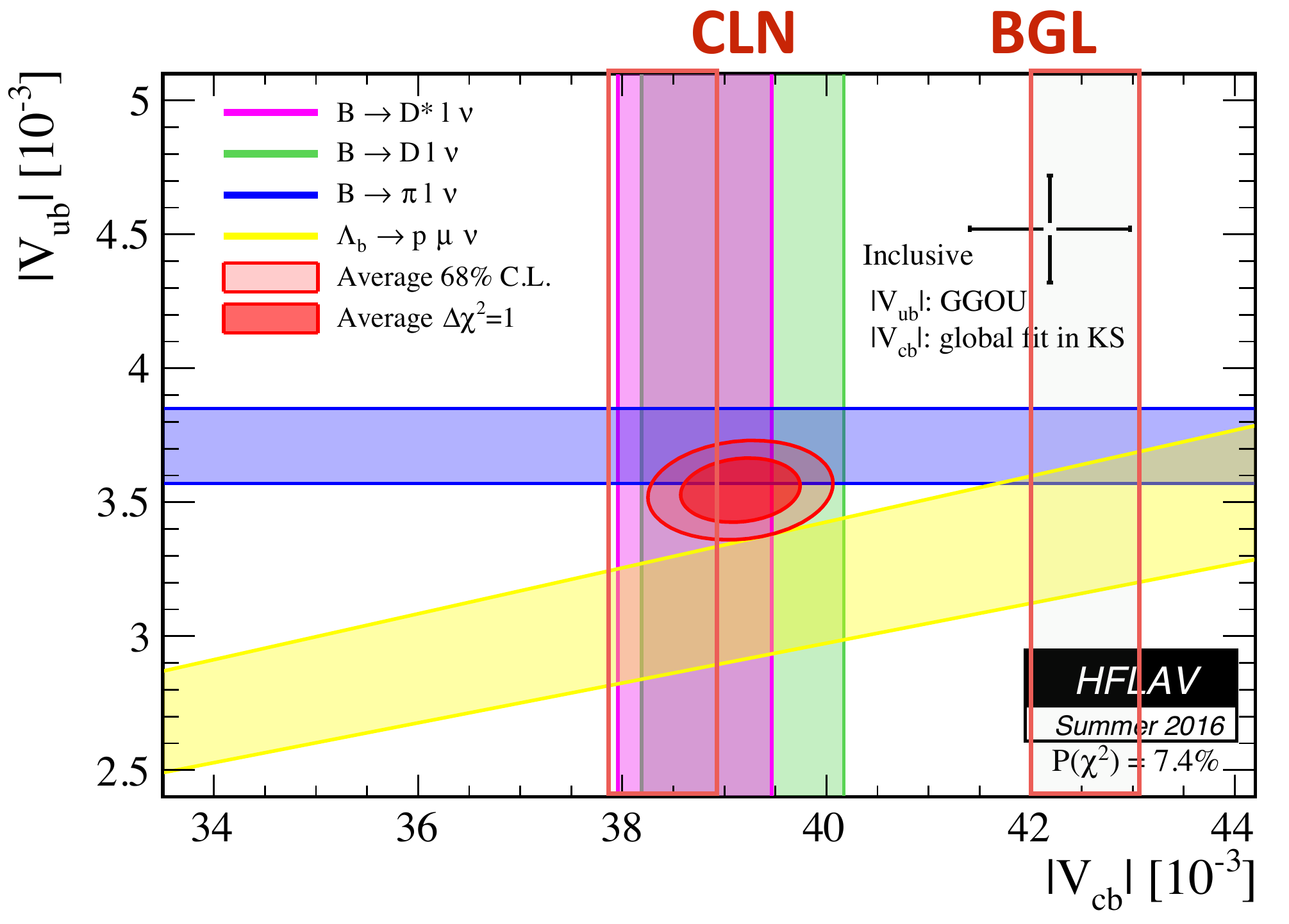}
\caption{Left: The $CP$ violation angle $\gamma\, (\phi_3)$~\cite{Rahatlou}. Right: determinations of $|V_{cb}|$ and $|V_{ub}|$~\cite{Waheed}.}
\label{unitarity}
\end{center}
\end{figure}

One of the most interesting new results is  the more precise determination of the unitarity triangle angle $\gamma\, (\phi_3)$,
dominated by a number of new and updated results from LHCb~\cite{Rahatlou,Urquijo} [Figure~\ref{unitarity}].
Another is that the long-standing discrepancy between the exclusive and inclusive determinations of $|V_{cb}|$
has been resolved by a new Belle determination from $B\rightarrow D^\ast \ell\nu$, analyzed in a  framework with minimal theoretical assumptions and utilizing measured form factors~\cite{Waheed}  [Figure~\ref{unitarity}]. Hopefully, similar progress will be made on the similar $|V_{ub}|$ discrepancy in the future.

\subsubsection*{Flavor anomalies}
The  flavor anomalies~\cite{Rahatlou,Tuning,Fajfer} in the  ratios
\[ R_{D^{(\ast)}}= \frac{BF (B\rightarrow D^{(\ast)} \tau\nu)}{BF (B\rightarrow D^{(\ast)} \mu\nu)}\quad  \text{and}  \quad R_{K^{(\ast)}}= \frac{BF (B\rightarrow K^{(\ast)} \mu^+\mu^-)}{BF (B\rightarrow K^{(\ast)} e^+e^-)}, \]
if confirmed, would imply a breakdown of lepton family universality. There is also a suggestion of new physics in the
$B\rightarrow K^\ast\mu^+\mu^-$ angular distributions, best described by an effective $V-A$ coupling  to the quarks and $V$ or $V-A$ to the leptons.

\begin{figure}[htbp]
\begin{center}
\includegraphics*[scale=0.5]{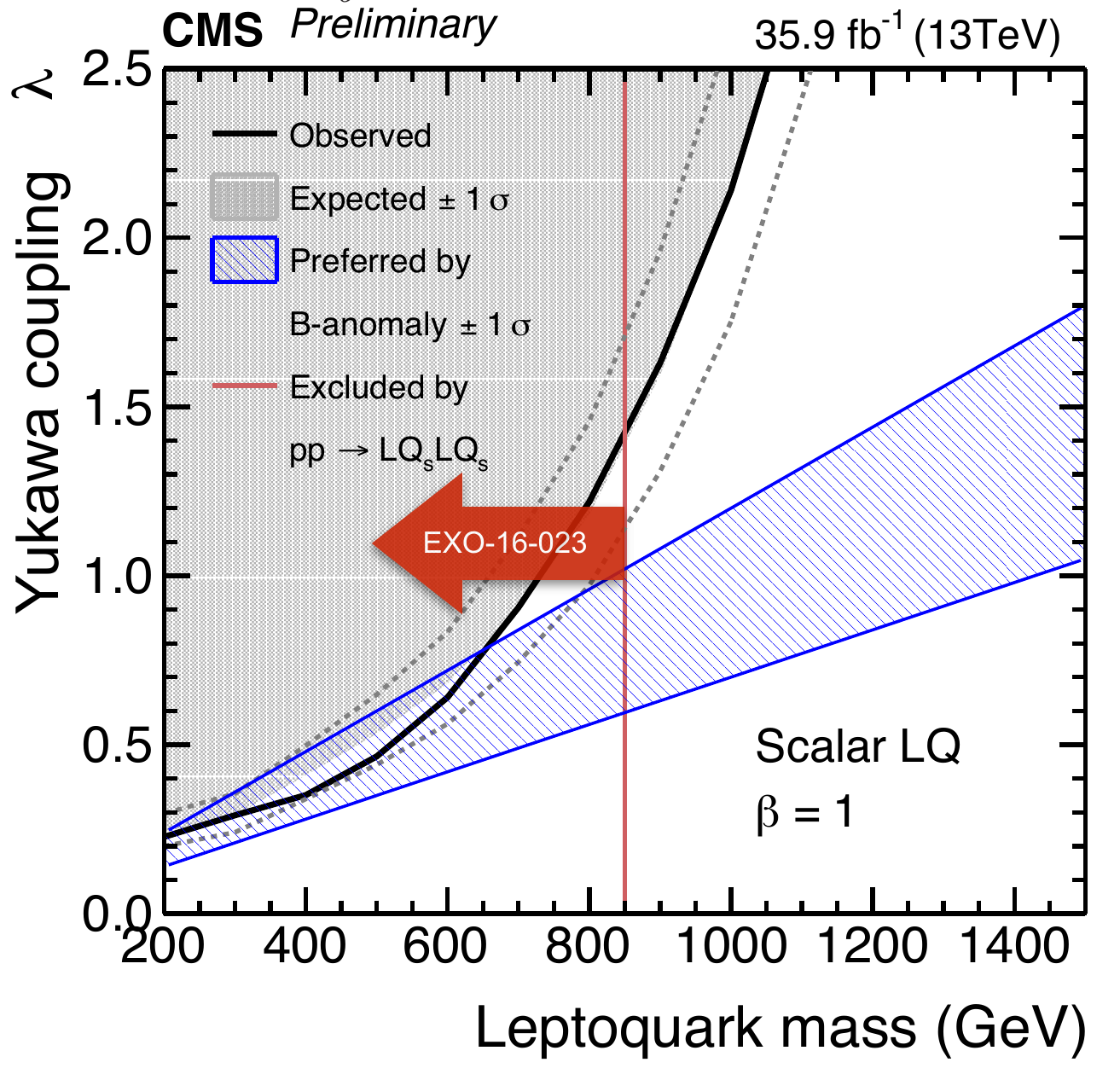}
\includegraphics*[scale=0.37]{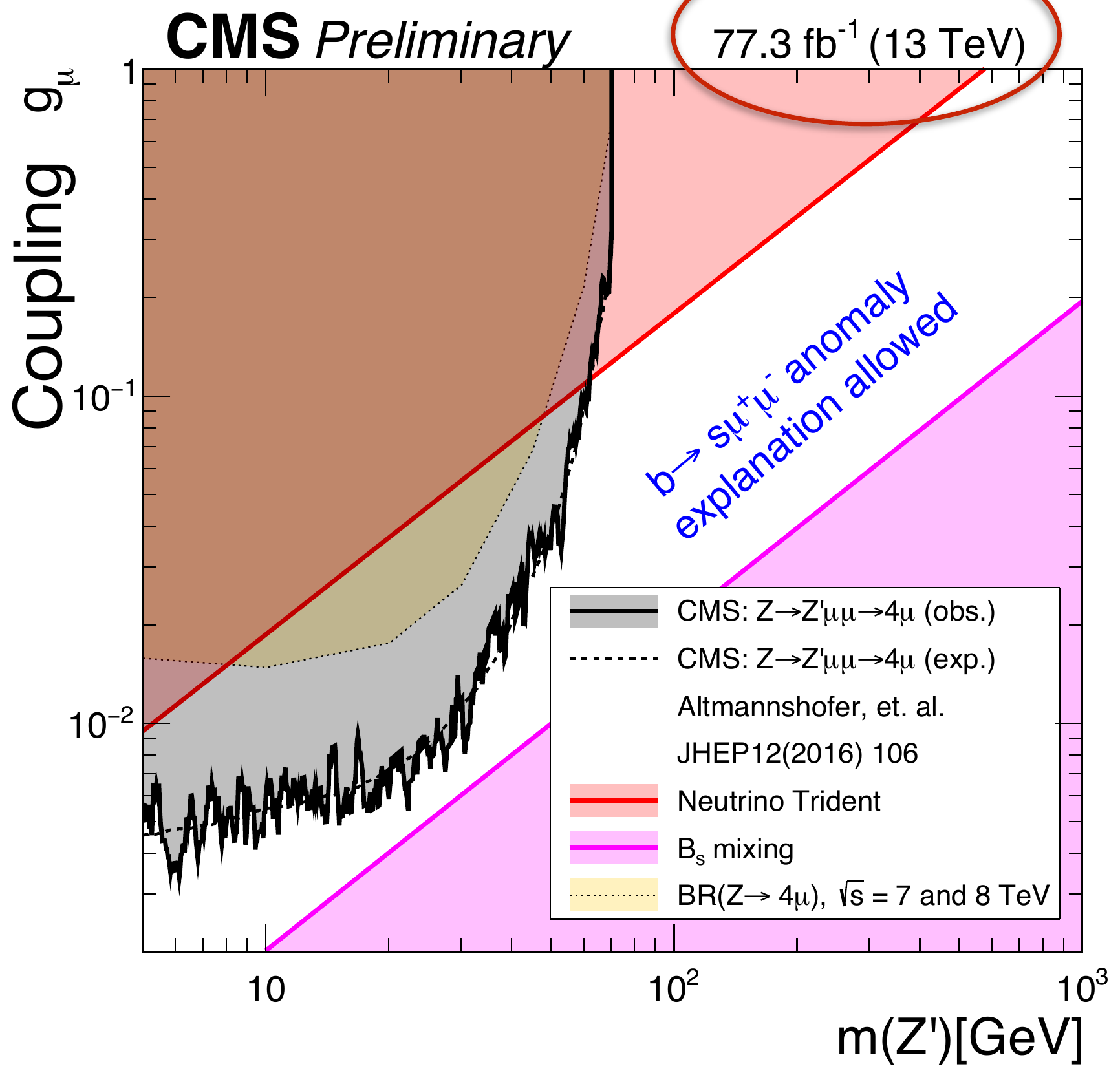}
\caption{Constraints on leptoquark (left) and $Z'$ (right) models suggested by the $B$ anomalies~\cite{Rahatlou}.}
\label{Banom}
\end{center}
\end{figure}

$R_{D^{(\ast)}}$ suggests a relatively large new physics effect since the SM decays occur at tree level, while new physics explanations for $R_{K^{(\ast)}}$ only have to compete with (small) SM penguins.
Many authors have suggested the possibility of a leptoquark to account for $R_{D^{(\ast)}}$ , while either a leptoquark or flavor-violating $Z'$ have been suggested for $R_{K^{(\ast)}}$. Constaints on some specific models are shown in Figure~\ref{Banom}.

\subsubsection*{Other important results, limits, prospects, or anomalies}
Many other important topics in flavor physics were discussed~\cite{Carli,Rahatlou,Tuning,Fajfer,Mihara}, including:
\begin{itemize}
\item $B$ and $D$ decays; $CP$ violation; and the CKM matrix.
\item $B_{s}\rightarrow \mu^+\mu^-$, $B\rightarrow \mu^+\mu^-$.
\item $K^+\rightarrow \pi^+\nu\bar\nu$,  $K^0\rightarrow \pi^0\nu\bar\nu$.
\item $\epsilon'/\epsilon$.
\item $g_\mu-2$
\item Lepton-flavor violation in $\mu\ra e\gamma$, $\mu\ra 3e$, $\mu\ra e $ conversion.
\item The muonic Lamb shift (aka the proton radius).
\item Electric dipole moments.
\item The neutron lifetime discrepancies.
\end{itemize}

\subsection{Strong interactions}
There has been tremendous progress in lattice QCD, with applications to
weak interaction decay  constants, form factors, and mixing; hadronic vacuum polarization and light-by-light
scattering relevant to $g_\mu-2$; neutrino, muon, and dark matter interactions with nucleons and nuclei;
and parton distribution functions~\cite{Shanahan}.

\begin{figure}[htbp]
\begin{center}
\begin{minipage}{5.5cm} 
\hspace{-1cm}
\includegraphics*[scale=0.35]{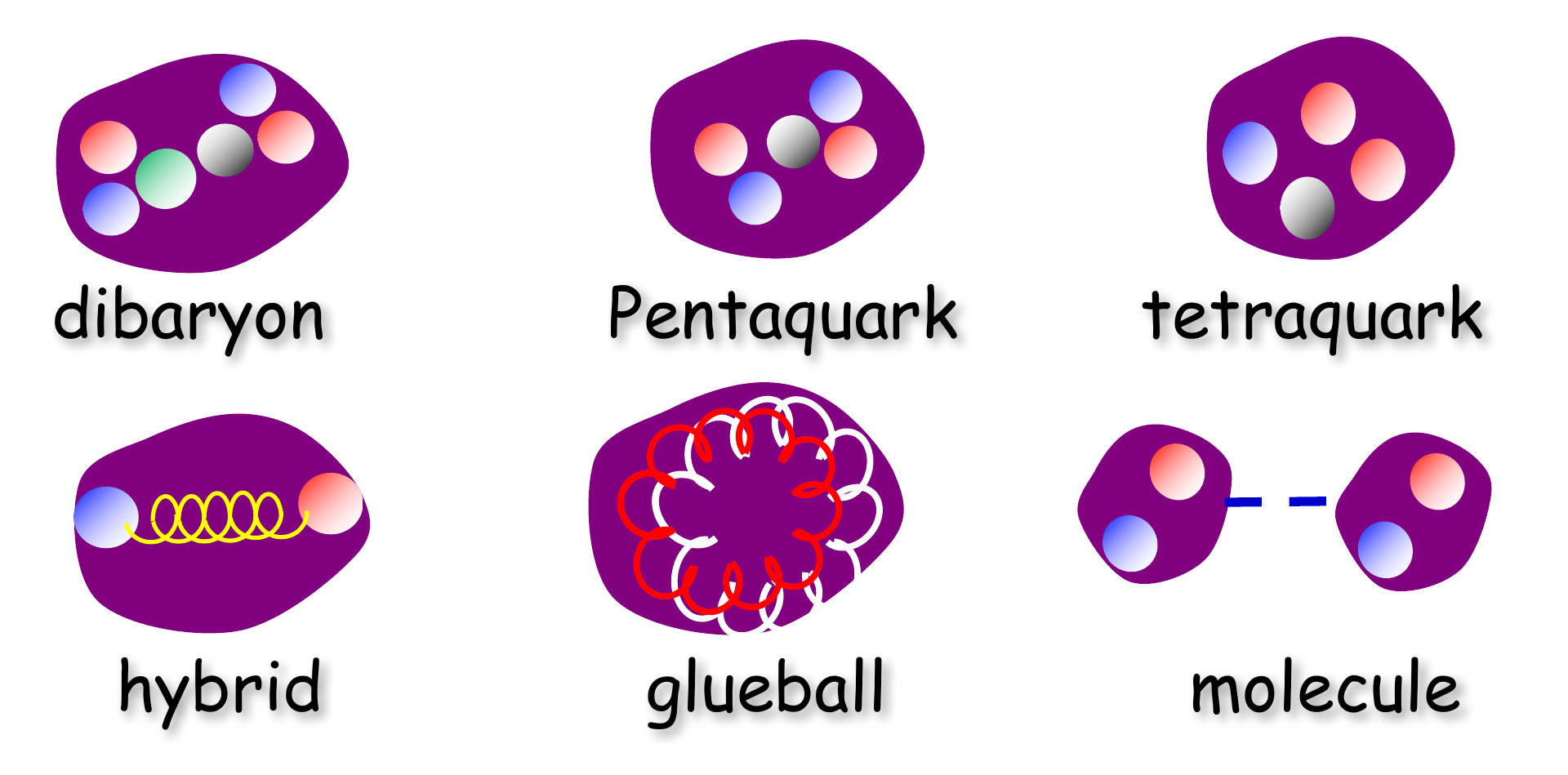}
\end{minipage}
\hspace*{1cm}
\begin{minipage}{5.0cm} 
\includegraphics*[scale=0.4]{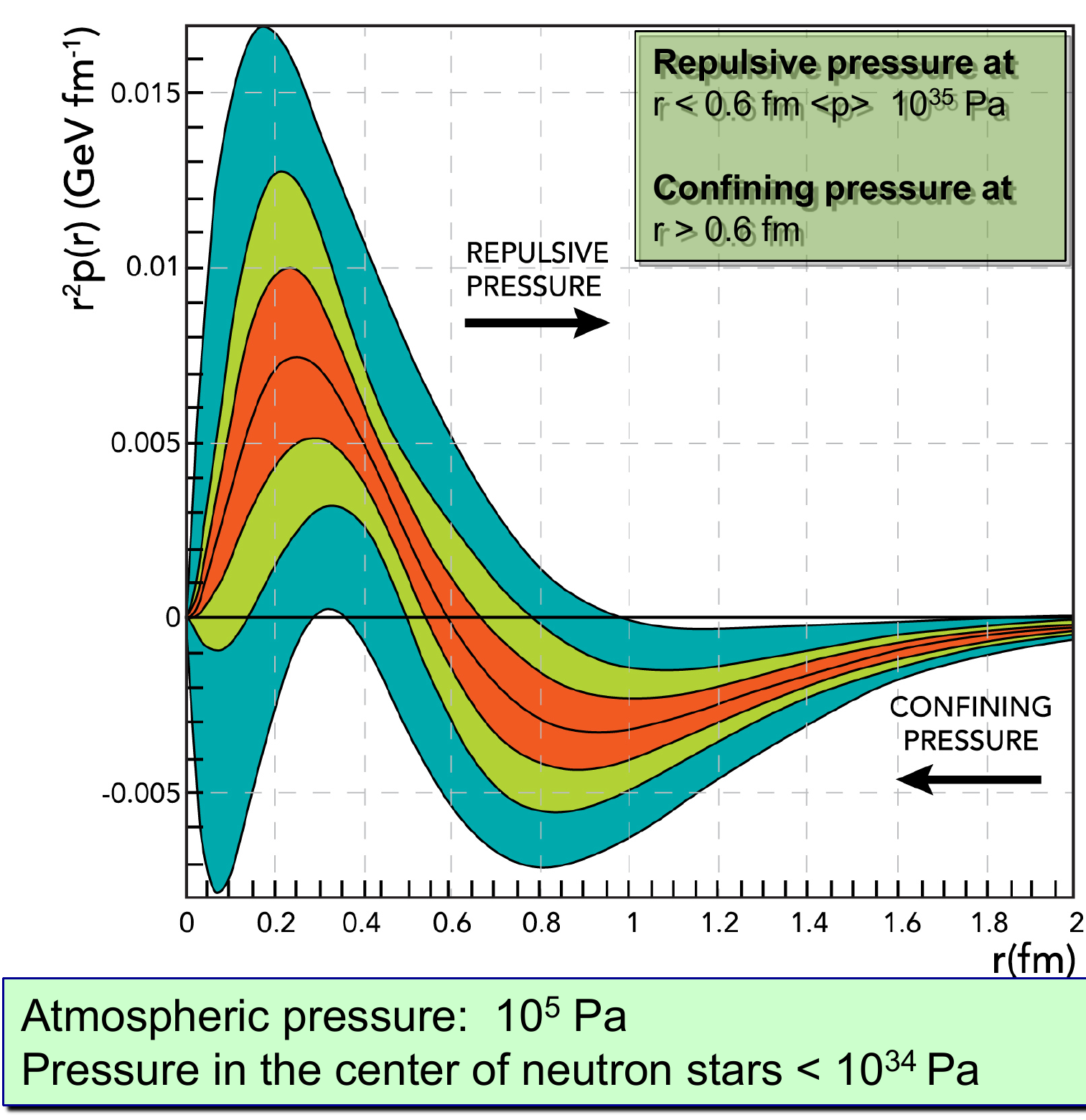}
\end{minipage}
\caption{Left: Exotic hadrons~\cite{Fang}. Right: Pressure distribution in the proton, obtained from GPDs~\cite{Elouadrhiri}.}
\label{strongints}
\end{center}
\end{figure}

There has been extensive experimental work on exotic hadrons, such as pentaquarks and tetraquarks. There are still many theoretical models, but  anticipated
experimental results should help clarify our understanding~\cite{Fang} [Figure~\ref{strongints}].

Experimental programs at Jefferson Lab ($e^-N, \gamma N$) and elsewhere are probing parton distributions, the spin structure, and other aspects of the nucleon. There has been a major effort to map the three-dimensional structure of the nucleon, as described by generalized parton distributions (GPDs)~\cite{Elouadrhiri} [Figure~\ref{strongints}].

\subsection{Neutrino physics}

Neutrinos are an important probe of (and are constrained by) particle physics, astrophysics, and cosmology.

\subsubsection*{The standard 3-neutrino picture}
\begin{figure}[htbp]
\begin{center}
\includegraphics*[scale=0.75]{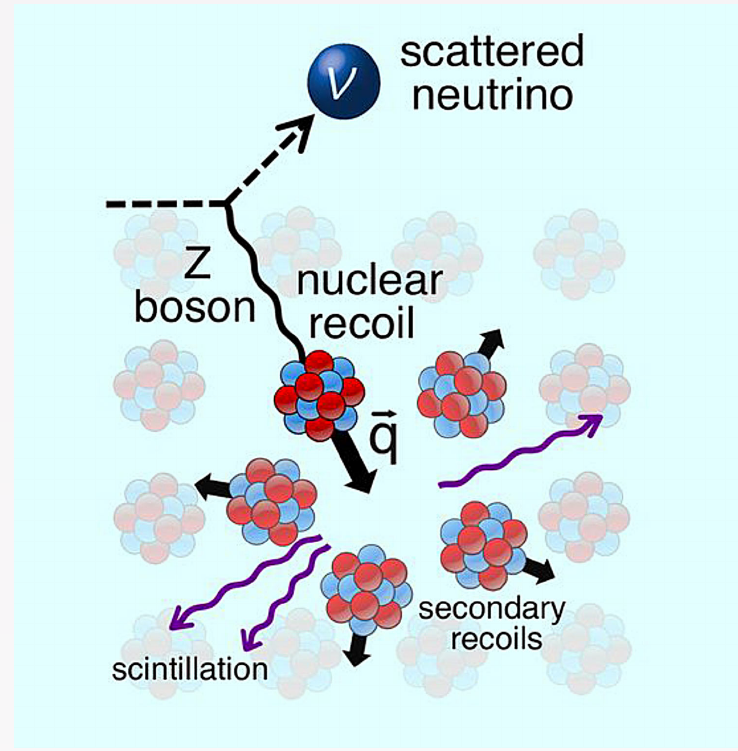}
\hspace{1cm}
\includegraphics*[scale=0.38]{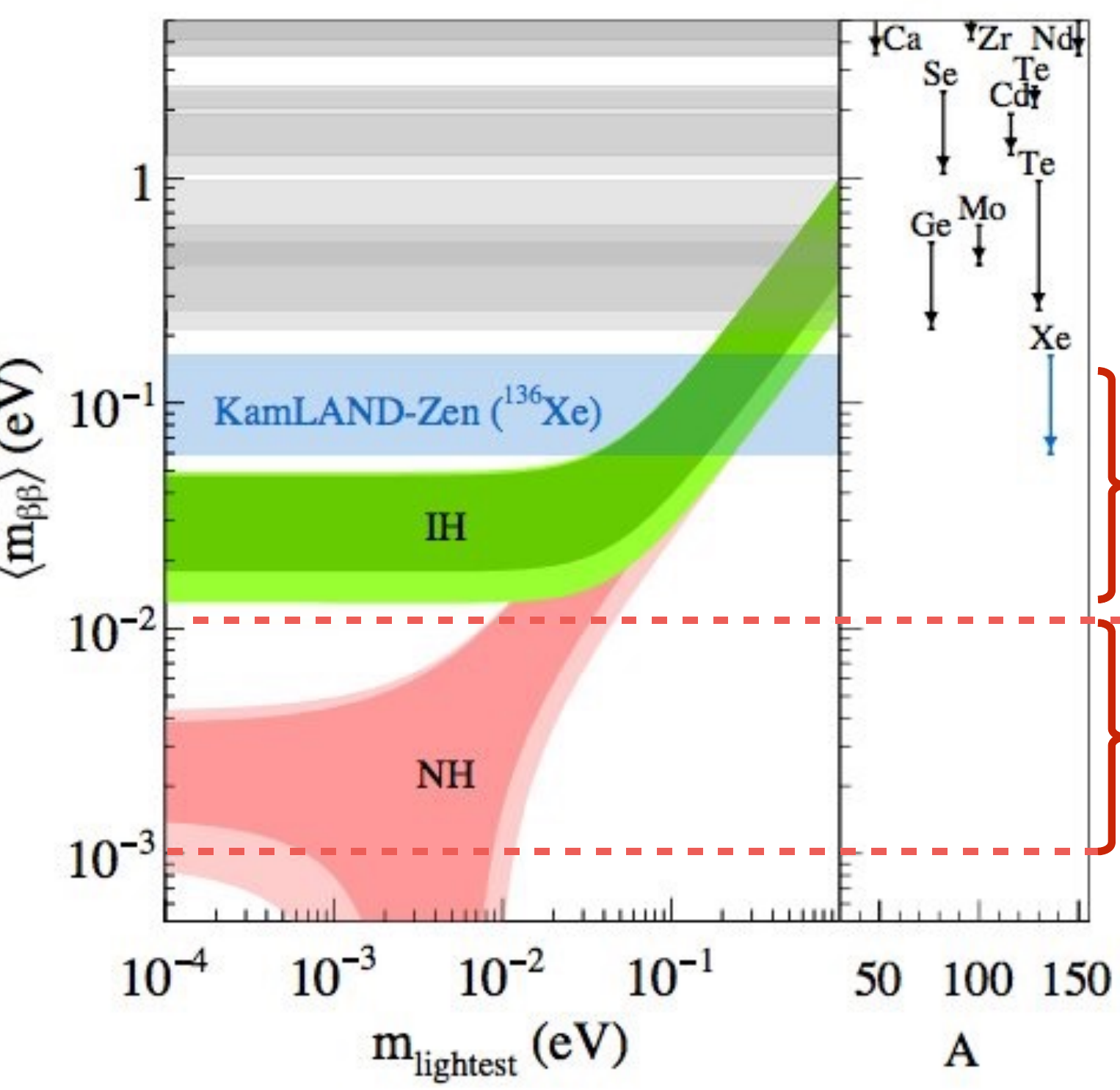}
\caption{Left: Coherent elastic $\nu$-nucleus scattering~\cite{Saakyan}. Right: ``Lobster-claw'' diagram of the effective  $\beta\beta_{0\nu}$ mass vs. the lightest mass~\cite{Saakyan}.}
\label{coherent}
\end{center}
\end{figure}

\begin{itemize}
\item Recent results within the 3-neutrino picture~\cite{Yokoyama,Petcov,Saakyan} include a measurement of $\theta_{13}$ from Double
Chooz, consistent with (but less precise than) previous Daya Bay and RENO results; final $\nu_\tau$ results from OPERA, with 10 candidate events; final results from MINOS/MINOS$+$; new T2K and NO$\nu$A  results (including the first  data from the NO$\nu$A $\bar\nu$ beam). The long baseline data indicate that the normal hierarchy is favored, a nonzero Dirac $CP$ violation phase is favored, and that the upper octant for $\theta_{23}$ is favored over maximal mixing or the lower octant~\cite{Yokoyama} [Figure~\ref{nupar}]. 

\begin{figure}[htbp]
\begin{center}
\includegraphics*[scale=0.2]{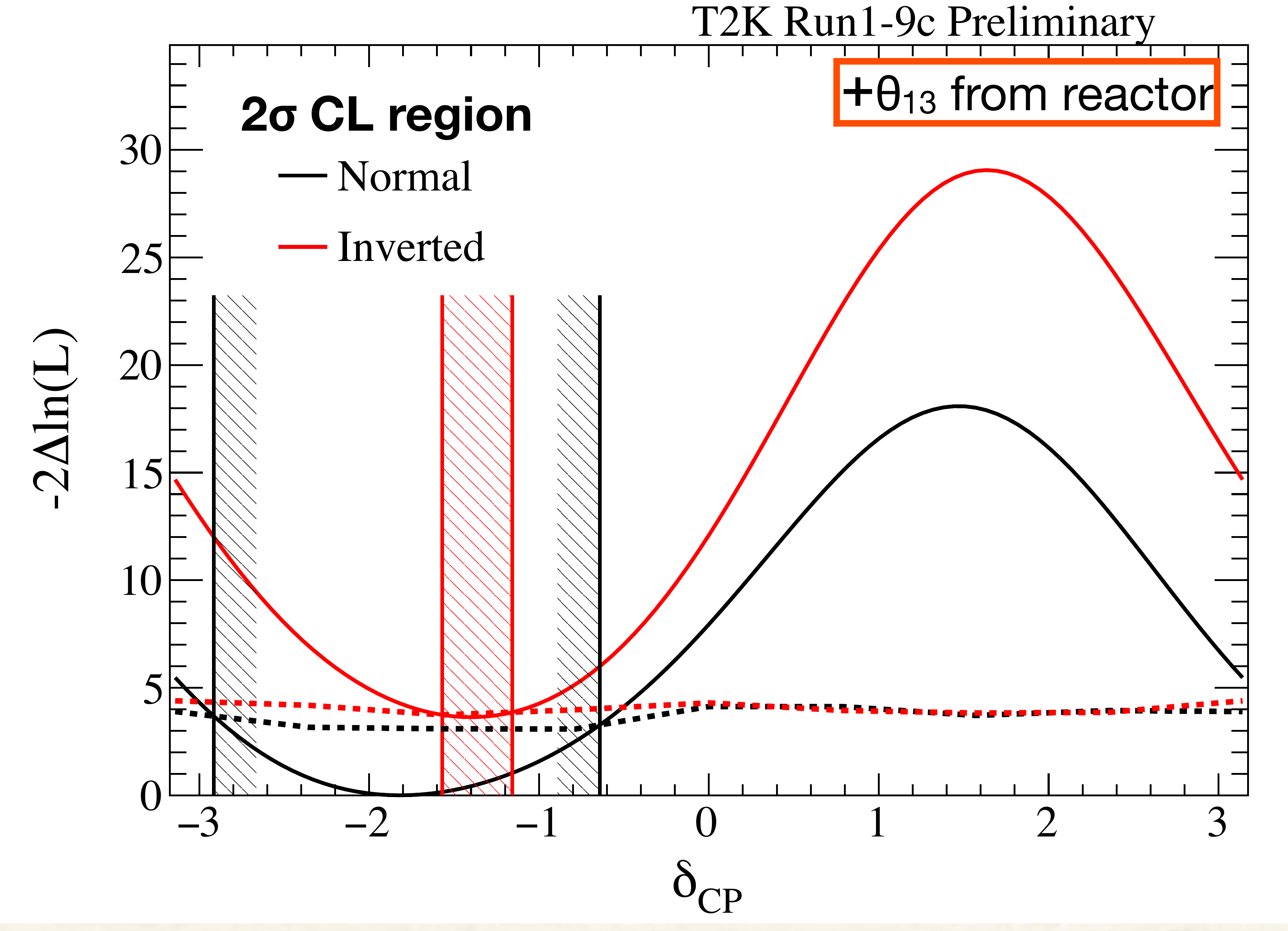}
\includegraphics*[scale=0.16]{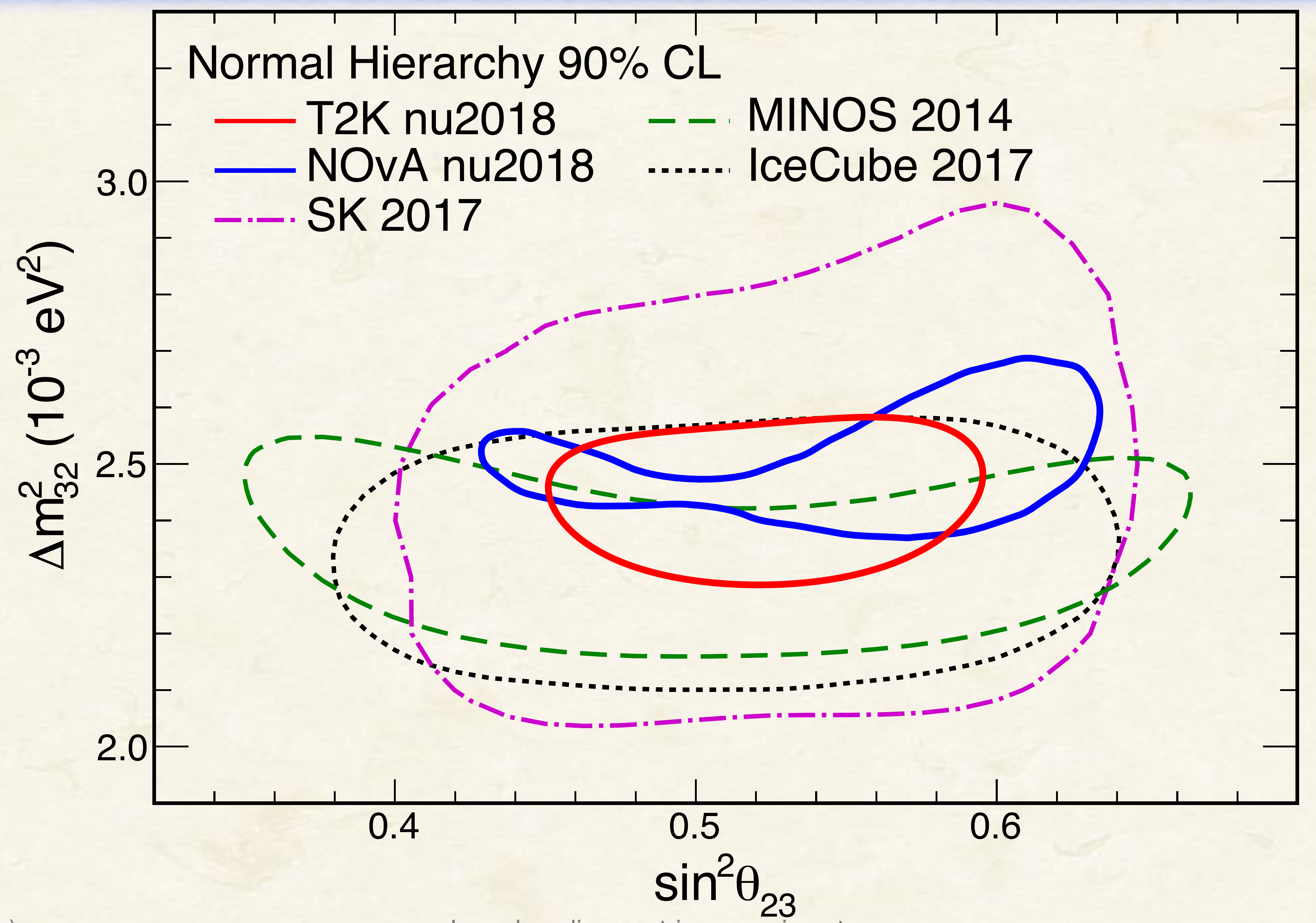}
\caption{Left: The Dirac neutrino $CP$ phase~\cite{Yokoyama}. Right: $\Delta m^2_{32}$ vs. $\sin^2 \theta_{23}$~\cite{Yokoyama}.}
\label{nupar}
\end{center}
\end{figure}

\item In addition to confirming the oscillation  results, it is still to be established whether the neutrinos are Majorana (e.g., as in some version of the seesaw model), with implications for leptogenesis, or Dirac (e.g., because the Yukawa couplings are forbidden to leading order by a new symmetry). Similarly, the observed mixings could be associated with  a new broken symmetry~\cite{Petcov}, or could be essentially anarchic (suggestive of  superstring theories).

\item Accurate measurements of low energy neutrino cross sections are essential for the interpretation of oscillation and other experiments. 
The Fermilab program includes new   MINER$\nu$A $\bar\nu$ data on a number of nuclei, and the Short Baseline Neutrino program (SBN), consisting of three liquid argon detectors: MicroBooNE (running), ICARUS, and SBND~\cite{Garcia}.

\item The COHERENT experiment has utilized the Oak Ridge Spallation Neutron Source to make the first measurement
of coherent elastic neutrino scattering from the entire nucleus~\cite{Saakyan} [Figure~\ref{coherent}].

\item  Neutrinosless double beta decay ($\beta\beta_{0\nu}$) is the only known practical way to probe the Majorana vs. Dirac nature of the light neutrinos. A number of experiments are now sensitive down to around 0.1 eV on the effective $\beta\beta_{0\nu}$
mass (with nonnegligible uncertainly from nuclear matrix elements). This is somewhat above the expected range for the inverted mass hierarchy, but excludes much of the degenerate region for Majorana neutrinos [Figure~\ref{coherent}]. The experiments should eventually
be senstive to around 0.01 eV, covering the entire inverted hierarchy region. Unfortunately,  probing the region expected for Majorana neutrinos with a normal hierarchy is probably impossible without novel or radical new ideas and a consolidated worldwide effort~\cite{Saakyan}.

\item The KATRIN trititum $\beta$ decay experiment in Karlsruhe has started running, with a projected sensitivity of 0.2 eV on the effective $\beta$ decay mass~\cite{Saakyan}.

\item Cosmological constraints on the sum of the active neutrino masses should eventually be senstive down to around 0.05 eV (the smallest value allowed by the oscillation experiments) or even better~\cite{Hazumi}.

\item Many other results were presented on Solar, supernova, atmospheric, high energy, relic, and geo neutrinos.

\end{itemize}

\subsubsection*{Possible sterile neutrinos}
Most models of neutrino mass involve sterile (aka right-handed or singlet) neutrinos, which may be light (as in Dirac neutrinos)
or very heavy (as in seesaw models). However, the sterile neutrino-induced oscillations suggested by LSND would require mixing between ordinary and sterile neutrinos of the same helicity, which is nontrivial to achieve. (It would require \emph{two} types of small mass terms, usually  Dirac and Majorana.)

\begin{figure}[htbp]
\begin{center}
\includegraphics*[scale=0.7]{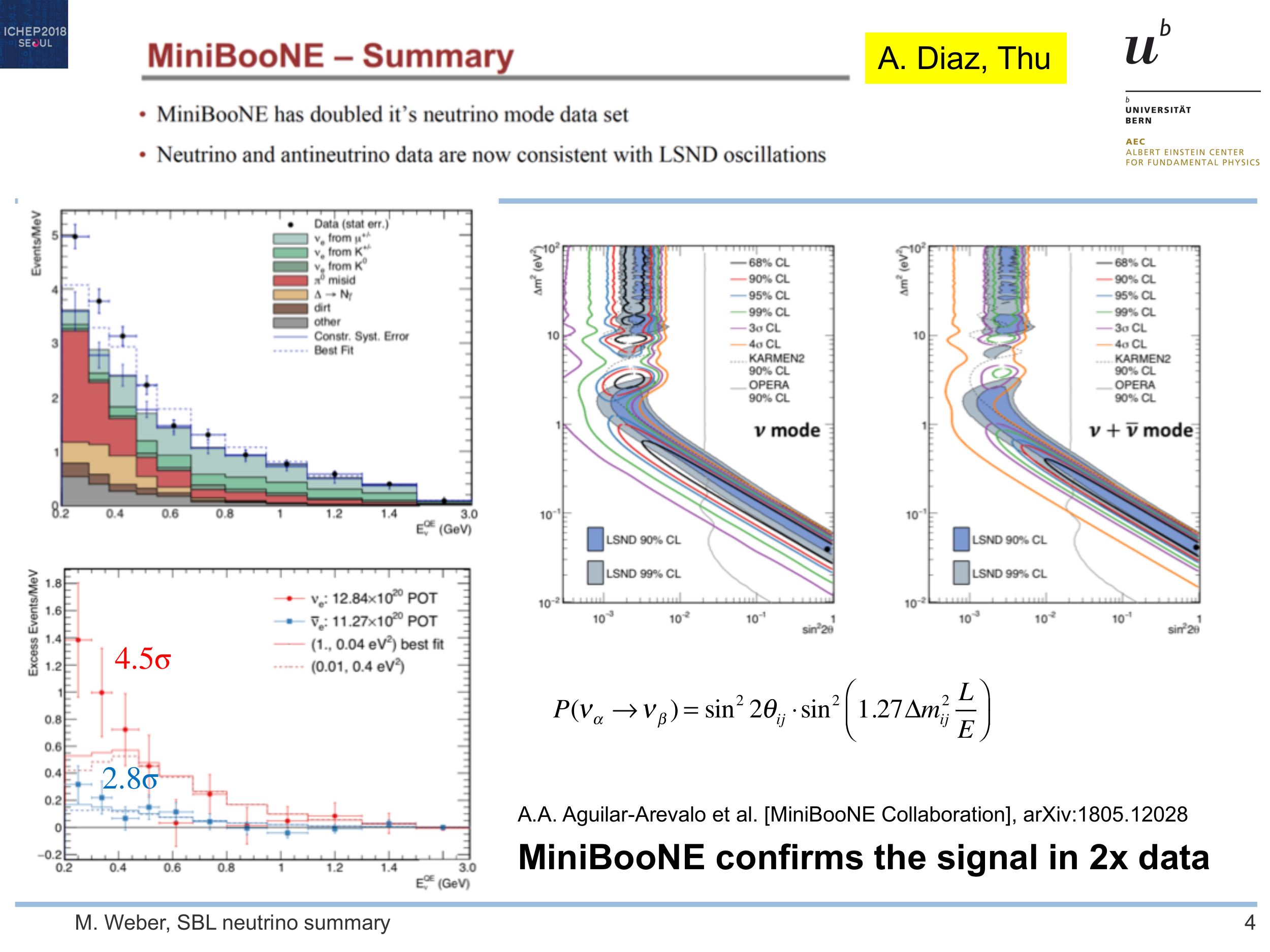}
\includegraphics*[scale=0.7]{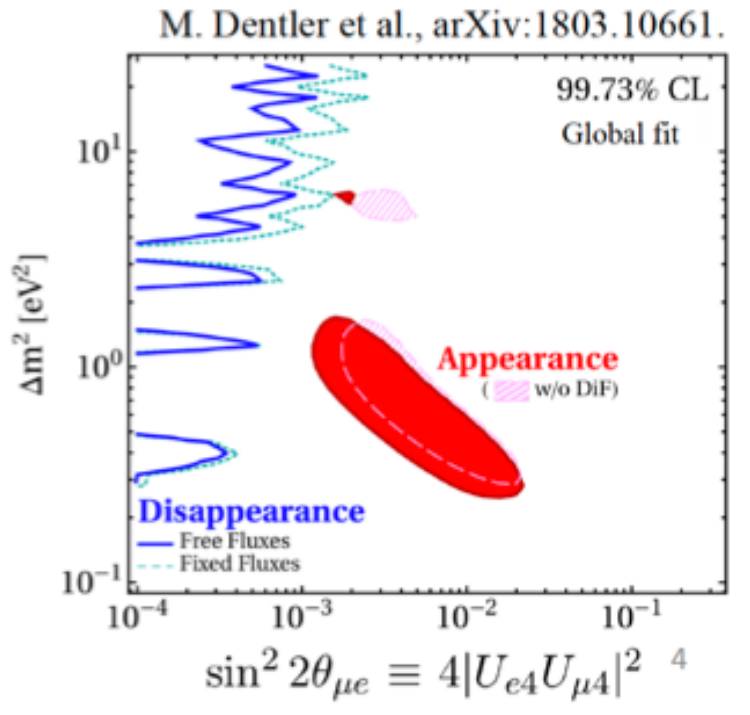}
\caption{Left: MiniBooNE spectrum~\cite{Weber}. Right: Sterile neutrino appearance (signal) and disappearance (exclusion) regions~\cite{Weber}.}
\label{sterile}
\end{center}
\end{figure}

MiniBooNE has doubled its neutrino data set. An oscillation fit to the neutrino and antineutrino data for $\snu_\mu \ra \snu_e$  yields results consistent with LSND, suggesting
 an $\sim$1 eV sterile neutrino\footnote{However,  the  low-energy $\nu_e$ data is not well described by the oscillation hypothesis [Figure~\ref{sterile}], emphasizing the importance of the cross section measurements.}~\cite{Weber} [Figure~\ref{sterile}]. On the other hand, the evidence for sterile neutrinos from the reactor anomaly is weakened by the Daya Bay and RENO measurements of the time dependence of the reactor fluxes, suggesting that the deficit of $\bar\nu_e$ is associated with those from $^{235}U$ but not $^{239}Pu$~\cite{Saakyan}. There is also a strong conflict with disappearance experiments  [Figure~\ref{sterile}]
 and with cosmological constraints (unless one invokes nonstandard cosmologies). There is currently no coherent picture of what is going on. Hopefully, the situation will soon be clarified by a new generation of short baseline accelerator (SBN),  reactor (DANSS, NEOS, PROSPECT, STEREO, SoLid), and source experiments~\cite{Weber}.

\subsection{Dark matter}
There are numerous candidates for dark matter~\cite{Drees}. These include primordial black holes (which are strongly constrained by astrophysics,
but may be possible in the mass range observed by LIGO) and many possibilities for particle dark matter. The latter include Weakly Interacting Massive Particles (WIMPs),  dark sector particles, light gravitinos (as in gauge mediation), WIMPZILLAs (extremely heavy particles, e.g., $10^{10}$ GeV), fuzzy dark matter [extremely light bosons with de Broglie wavelength $\mathcal{O}$(kpc)],  warm dark matter (e.g., keV-scale sterile neutrinos),  moduli,  and axions or axion-like particles (ALPs). Dark matter could also be self-interacting or involve two or more components.
Another (non-dark matter) possibility is Modification of Newtonian Gravity (MOND),
which, however, has difficulty with large scale effects and the Bullet cluster.

\begin{figure}[htbp]
\begin{center}
\includegraphics*[scale=0.41]{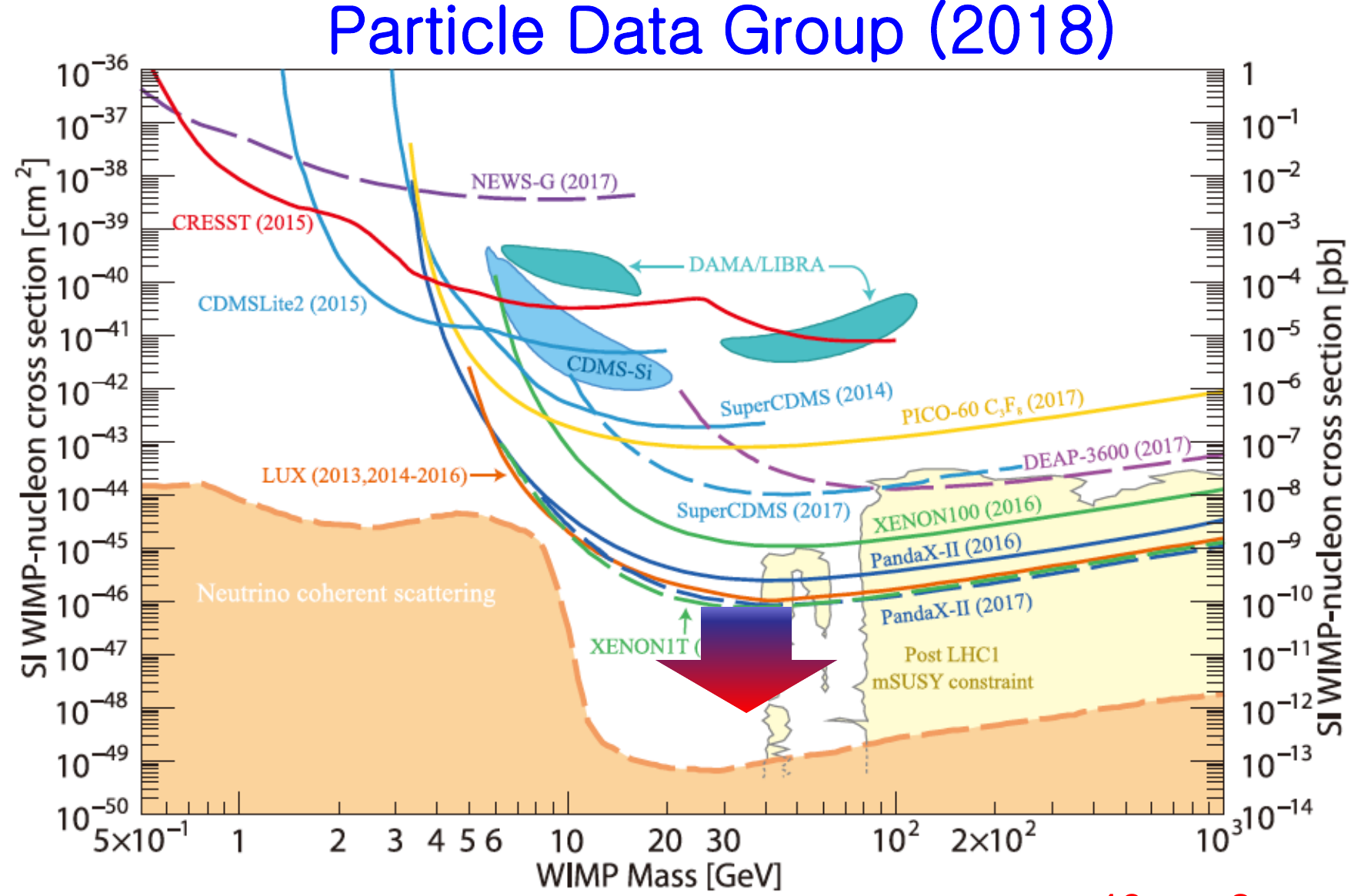}
\includegraphics*[scale=0.28]{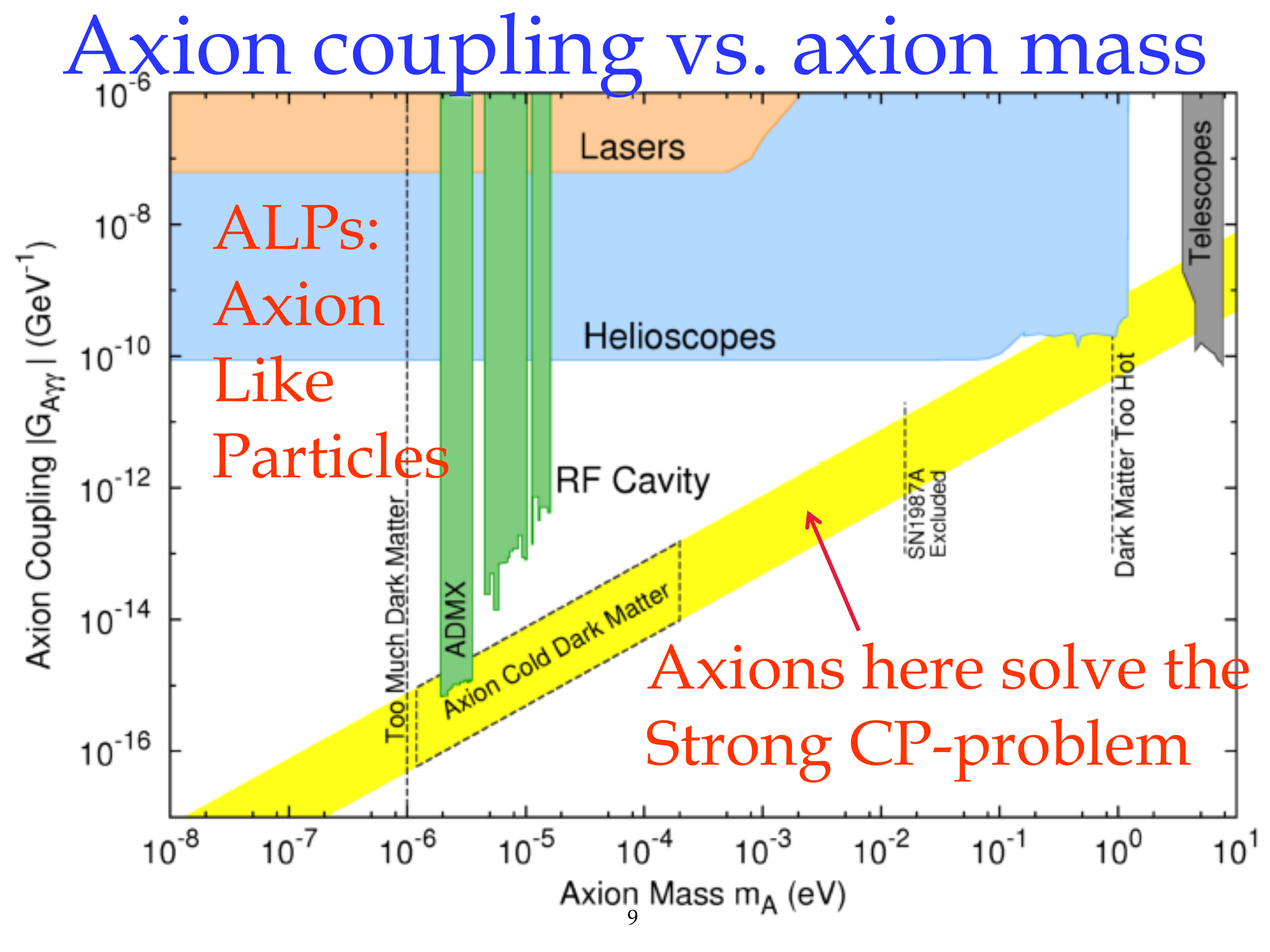}
\caption{Left: Limits on spin-independent WIMPs. Also shown are the neutrino floor and the allowed region from a minimal supersymmetric model~\cite{Lee}. Right: Constraints on axions~\cite{Semertzidis}.}
\label{dark}
\end{center}
\end{figure}

There have been numerous direct~\cite{Lee} (interacting with target) and indirect~\cite{Kounine} (observation of annihilation or decay products)
searches for dark matter [Figure~\ref{dark}], with no confirmed signal.\footnote{The DAMA/LIBRA signal from an annual modulation is statistically overwhelming (12.9$\sigma$), but other experiments  exclude a standard WIMP interpretation. Clarification is greatly desired.} The direct detection WIMP experiments are approaching the neutrino background floor (from Solar and atmospheric neutrinos), which will limit further progress without dedicated techniques such as directionality. Very light WIMPs are also possible. Possible loopholes in the interpretation include a local variation in the dark matter density and multi-component dark matter.

There have also been extensive searches at the LHC for the production of (unobserved except as missing transverse energy) dark matter particles in association with a jet, $\gamma$, $W$, $Z$, Higgs, \ldots~\cite{delray}.

Axions are moltivated not only by the strong $CP$ problem but also by string theory. The ADMX cavity experiment is
starting to probe the strong $CP$ parameter region, and future experiments are expected to reach most of the interesting region~\cite{Semertzidis} [Figure~\ref{dark}]. 

\subsection{Cosmic rays}
\begin{figure}[htbp]
\begin{center}
\includegraphics*[scale=0.3]{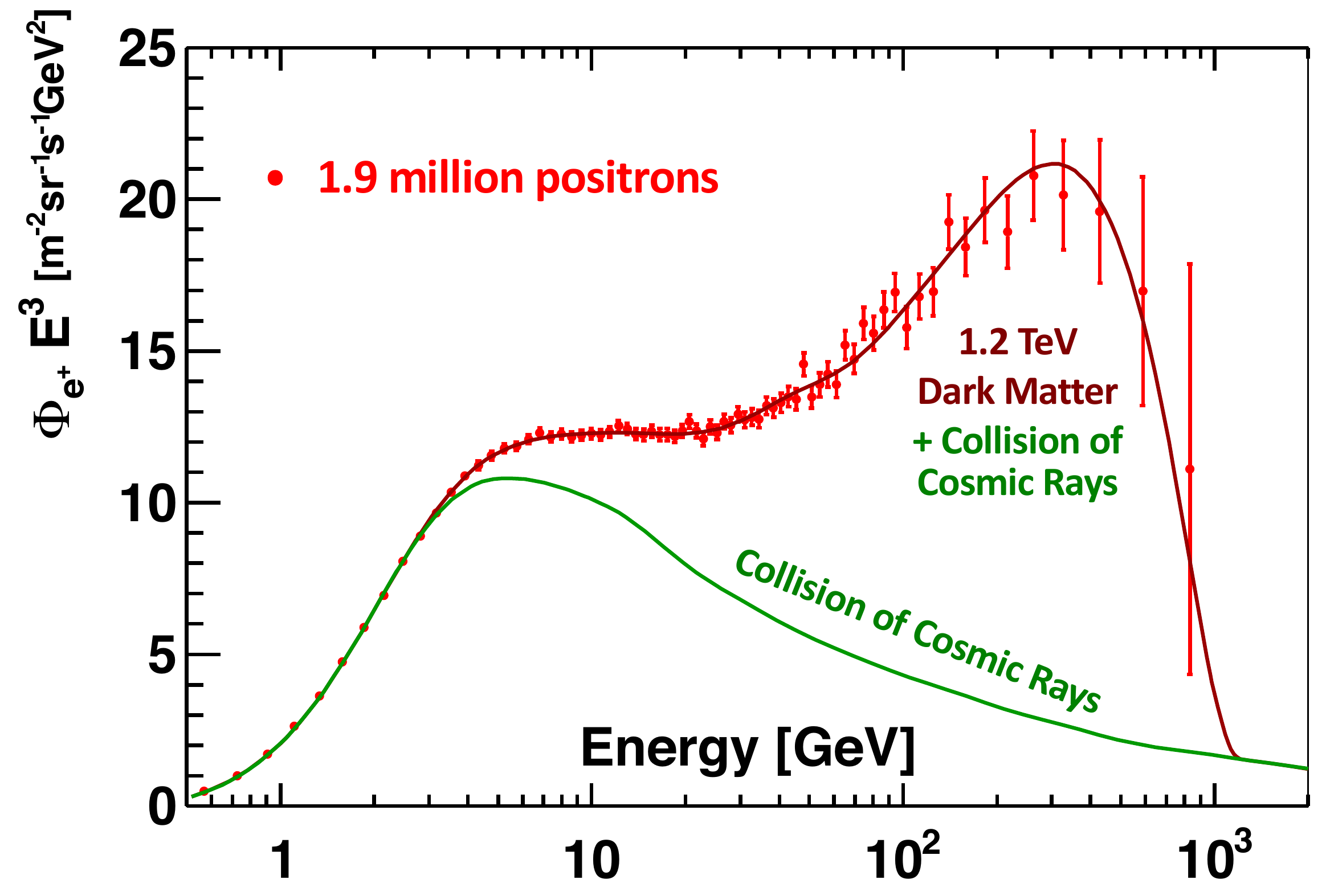}
\includegraphics*[scale=0.3]{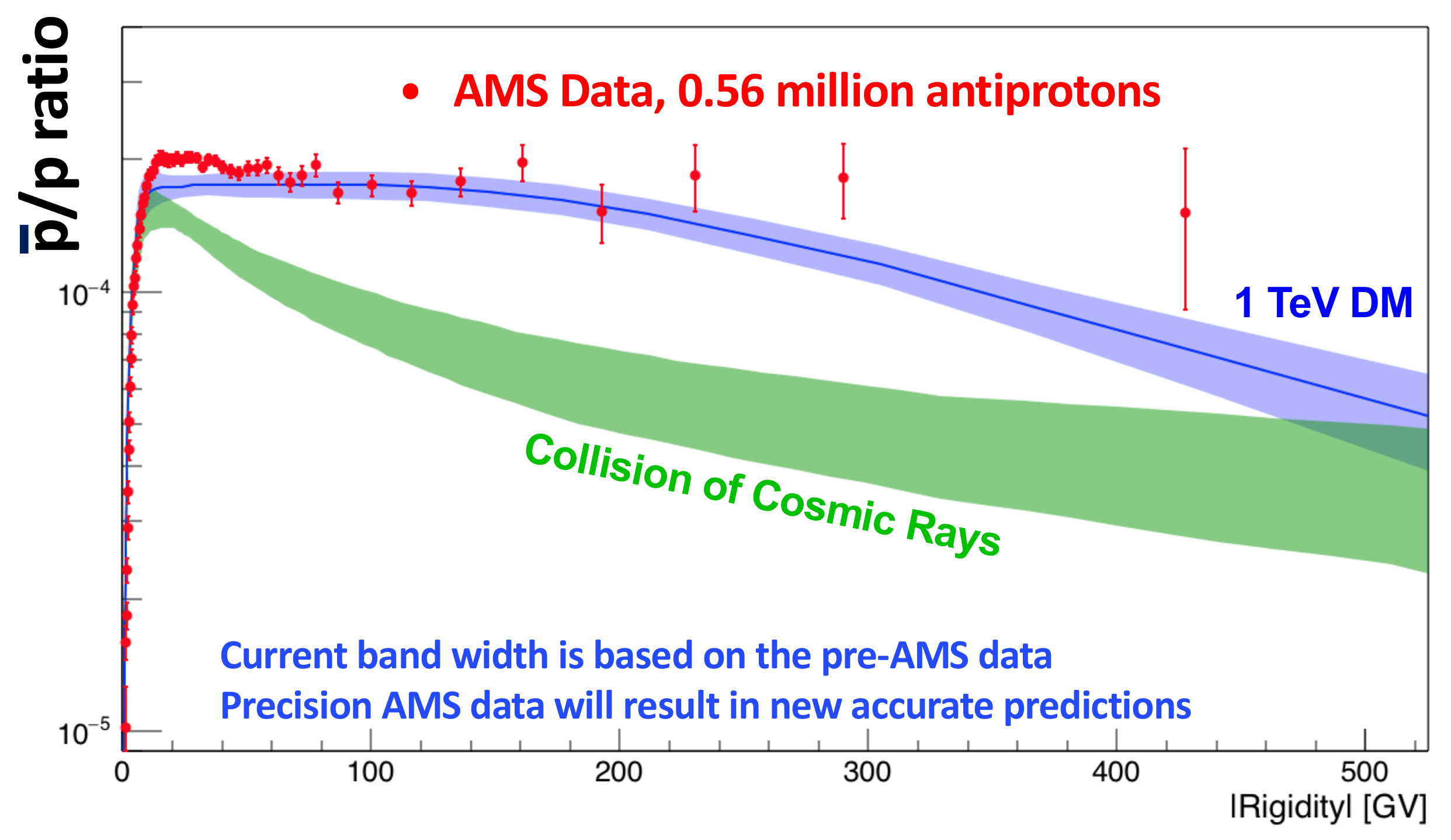}
\caption{Recent AMS observations of $e^+$ and $\bar p$~\cite{Kounine}.}
\label{cosmic}
\end{center}
\end{figure}

There has been extensive observational work on high energy cosmic rays, up to some $10^{12}$ GeV/particle~\cite{Kounine}.
Much of this is focussed on astrophysical issues, such as their origin and composition. More relevant for particle physics
are the fluxes of $e^+$, $\bar p$ and other antiparticles, which could be produced by dark matter anihilation, by the collisions of ordinary cosmic rays with interstellar matter, or from such astrophysical sources as pulsars.
Recent precise data on high energy $e^+$ and $\bar p$ from the Alpha Magnetic Spectrometer (AMS) 
indicates considerably more antiparticles than predicted by previous estimates of the secondary production rate, but is consistent with the annihilation of TeV-scale dark matter~\cite{Kounine}  [Figure~\ref{cosmic}]. 

 High energy neutrinos have been extensively studied by the
 IceCube collaboration.\footnote{Note added: after the conference IceCube announced the observation of a 0.29 PeV event coincident in direction with a known $\gamma$-ray blazar, initiating an extensive multi-wavelength observation program~\cite{Klein:2018waq}.} They have recently observed two double cascade events~\cite{Dujmovic}, consistent with $\nu_\tau$.
This would  imply a neutrino flavor ratio $\nu_e:\nu_\mu:\nu_\tau\sim 1.05:1.35:0.6$, consistent with  $1:1:1$ expected from a pion decay source after taking oscillations into account. 

\subsection{Gravity waves}
\begin{figure}[htbp]
\begin{center}
\includegraphics*[scale=0.4]{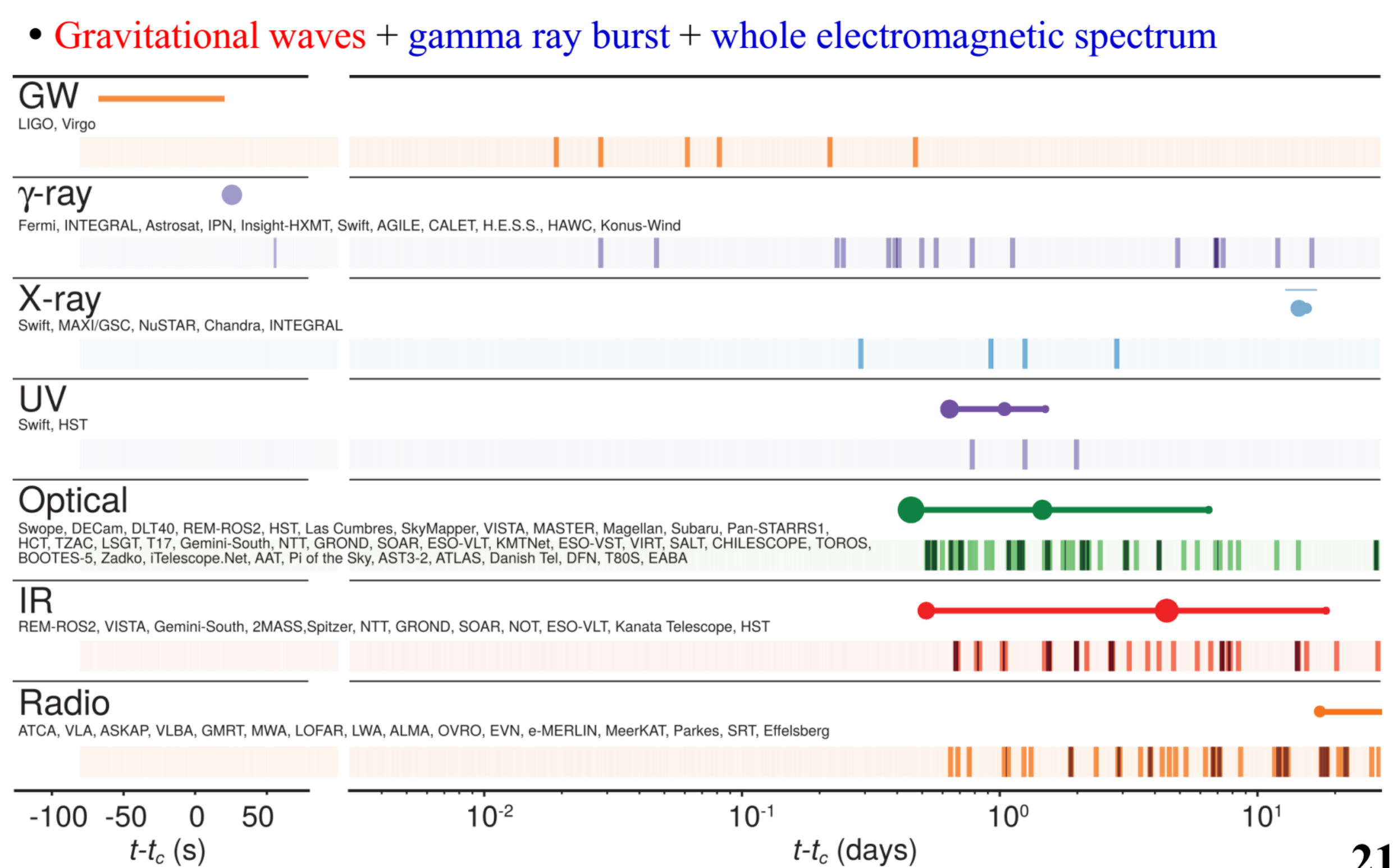}
\caption{Multi-messenger observations from GW170817~\cite{Shawhan}.}
\label{gravity}
\end{center}
\end{figure}

The LIGO and Virgo interferometers have initiated a whole new field in astrophysics, general relativity, and particle physics.
The observed binary  black hole (BH-BH) mergers involved black holes in the $\mathcal O (10 M_\odot)$ range and tested GR, e.g.,
setting an upper  limit of $1.2 \x 10^{-23}$ eV on the graviton mass~\cite{TheLIGOScientific:2016src}.
The neutron star (NS-NS) merger GW170817 led to a major multi-messenger observation program including a $\gamma$
ray burst and multiple electromagnetic wavelengths~\cite{Shawhan} [Figure~\ref{gravity}]. Implications include the possibility of using such mergers as a standard siren; that they are a site for  the $r$-process for the synthesis of elements heavier than $^{56}Fe$; that the velocity of a gravitational wave relative to light is constrained by $-3\x 10^{-15} < \beta_G-1< 7\x 10^{-16}$;	
and that some dark energy models are excluded.

Future gravitational wave results are expected to have particle physics implications for cosmological phase transitions, topological defects, and inflation.

\subsection{Other topics}
Many other important topics were discussed, including:
\begin{itemize}
\item Formal theory~\cite{Takayanagi}.
\item Astro-particle topics such as the  CMB, dark energy, and inflation~\cite{Hazumi}.
\item Future facilities~\cite{Akai,Lou,Yu}.
\item Detector and accelerator development~\cite{Bortoletto}.
\item Computing~\cite{Sexton}.
\item Technology applications and industrial ties~\cite{Riedler}.
\item Education and outreach~\cite{Shaw}.
\item Diversity~\cite{Yacoob}.
\end{itemize}

\section{Thoughts for the future}
I often hear  my colleagues express frustration about the recent lack of exciting new discoveries.
However, it is sometimes useful to step back and take a longer view. For example,  Newtonian gravity was spectacularly successful. It survived intact for some 230 years before being supplanted by general relativity (or at least 175 years until the deviation in Mercury's perihelion was established). 
In contrast, the Standard Model is equally successful, but is scarcely 50 years old (and the Higgs discovery was only six years ago). If the SM had fallen quickly and easily it would not be such a great accomplishment.
We should take pride in the SM and also the Standard Cosmological Model, which was developed in parallel. They will figure prominently in the history of science.

Of course, as human beings with relatively short lifetimes we have both psychological (i.e.,~our enthusiasm for what we are doing) and practical (e.g., funding, jobs, interest in the field by the public and especially by young potential scientists) reasons to hope for new developments. Fortunately, despite the SM successes there are strong reasons (Section~\ref{smsummary}) to believe that there is much still to be learned. 
Perhaps there will indeed be new layer(s) of physics at the (multi-)TeV scale. Or perhaps we may eventually be able to
construct a New Standard Model of Nature that incorporates cosmology and extends all the way  to the Planck scale.
I am optimistic that one or both of these will occur. The time scale is less certain. Next week? In 20 years? In 100 years?

\subsection{Ideas (conventional and not) for new physics}

There are many ideas for possible BSM physics, including:
\begin{itemize}
\item Symmetries, such as supersymmetry or family symmetries.
\item Compositeness or other strong dynamics, e.g., composite Higgs, dynamical symmetry breaking, or composite fermions.
\item Extra dimensions, which could be ``large'' (compared to the Planck length of $\sim 10^{-33}$ cm) and/or warped.
\item Dark or (quasi-)hidden sectors. These could be associated with dark matter, supersymmetry-breaking, or be ``random''.
\item Unification, such as grand unification, superstring theories, AdS/CFT, or holography.
\item New dynamical ideas, such a relaxion models, nnaturalness, clockwork models, or string instantons.
\item 
Random or environmental selection, such as in the multiverse or variations.
\item Remnants, i.e., new physics that does not solve specific SM problems, but emerges more or less accidentally
from an underlying  theory. Typical superstring theory remnants include $Z'$s, vector fermions, extended Higgs sectors, dark sectors, moduli, and axions~\cite{Halverson:2018xge}.
\item Here be dragons. Novel concepts, such as emergent interactions, dimensions, or spacetime (e.g., from quantum entanglement~\cite{Takayanagi}); hidden variables; \ldots.
\end{itemize}

\subsection{The paradigms}

It is usually assumed that Nature (except possibly for gravity) is described by a unique quantum field theory.
However, there are an infinite number of such theories, differing by the types of interactions,  fields, symmetries, couplings, and masses. How does Nature choose between them? Principles such as elegance, naturalness,  minimality, or ``uniqueness'' (within some ad hoc framework) are often explicitly or tacitly invoked,
but none of these are unambiguous or compelling.

Another issue is that Nature seems to be ``just right''~\cite{Schellekens:2013bpa}.
For example, the dark energy is small enough in magnitude for structures like galaxies to form before matter becomes too diffuse or space recollapses. Similarly, the interactions and parameters of the SM seem to be fortuitously arranged to allow for the existence of multiple types of stable atoms, and for stars, etc. Yet another example is that stable planetary orbits are only possible in three large space dimensions.

Moreover, these two issues, i.e.,  the selection principle on the infinite landscape of field theories, and the fact that
the one chosen is so suitable for our existence, must be satisfied simultaneously.

These types of questions take us somewhat out of the realm of the traditional scientific method into that of philosophy. In my opinion, however, developments in superstring theory and cosmology, and the absence (so far) of evidence for new physics relevant to the Higgs hierarchy problem and other SM shortcomings,  suggest that the time is ripe to examine  some of the usual paradigms~(e.g.~\cite{langacker:pup,Halverson:2018xge}).

\subsubsection*{Minimality or remnants}
Minimality, closely related to Occam's Razor, refers to the ansatz that any new physics should be as simple as possible and that it should solve one, or preferably more than one, SM problem. 

Minimality
 is often a good guide, but not always so. The human brain, DNA, and the Standard Model
do not appear to be minimal, and new physics might not be either. In particular, short distance theories such as superstrings, grand unification, or strong coupling compositeness  can easily lead to  features such as vector fermions or $Z'$s that
 ``slip through the cracks'' and remain light. These remnants might not solve any SM problems and are usually not minimal.

\subsubsection*{Naturalness or tuning}
Naturalness has frequently been a  good guide as to the existence and mass scales of new physics that could resolve
fine-tuning problems. Notable examples~\cite{Giudice:2008bi} include the electron mass,  $m_{\pi^+}-m_{\pi^0}$,  and $m_{K_L}-m_{K_S}$, which led (or could have led) to predictions of the positron, the $\rho$ (or constituent quark), and the $c$ quark and their mass scales. Similarly, the Higgs hierarchy problem apparently requires an unnatural fine-tuning by some 34 orders of magnitude in the SM, but can be cured if there is new TeV-scale physics such as supersymmetry, a composite Higgs, or large and/or warped extra dimensions. So far, however, ATLAS and CMS have not observed any sign of such effects. Even more telling is that the observed dark energy suggests a cosmological constant tuned by some 120 orders of magnitude compared to the Planck scale.  Perhaps naturalness has failed.

A plausible alternative to naturalness is the environmental (or anthropic) explanation of fine-tuning, i.e., that
it is required for Nature to be just right. 
This would make no sense if there were  a unique physics selected by some non-environmental mechanism. However, it is perfectly reasonable if it is a consequence of a very large landscape of vacua (such as are believed to exist in string theory) and some mechanism, such as eternal inflation, to sample them~\cite{Linde:2015edk}.
Structures such as galaxies, stars, and people would then emerge only in the suitable vacua.
In fact, environmental selection is probably the most plausible explanation of the small dark energy~\cite{Weinberg:1988cp}.
These issues will be further discussed in  the next section.

Other clever dynamical mechanisms, such as clockwork theories or relaxions, have been suggested as technically natural
solutions to  the fine-tuning problems.

\subsubsection*{Uniqueness or environment}\label{environment}
The uniqueness paradigm assumes that there is a unique and perhaps simple underlying theory of Nature. As commented above there are actually an infinite number of possible theories and no clear mechanism to understand which one
Nature chooses. Gauge theories offer a partial solution, in that the form of the interactions is constrained up to the group,
representations, coupling constant, and spontaneous symmetry breaking. However, Yukawa and scalar interactions are unconstrained unless new symmetries or principles are invoked.

There is a famous example of the failure of the uniqueness paradigm: the Mysterium Cosmographicum,
in which Johannes Kepler attempted to describe the relative radii of the six known planetary orbits by nesting the Platonic solids. Kepler's theory was reasonably successful given the accuracy of the observations at the time, but we now know that this was fortuitous. 
There are numerous planetary systems, and their orbits
are determined by  initial conditions and not uniqueness. Moreover, the lucky coincidence that the Earth has favorable conditions for life is not really a coincidence -- life can only develop where conditions are suitable.

The environmental paradigm is analogous to the modern understanding of the planetary orbit problem. It suggests that there may be no simple and unique explanation for some (or all?) of the features and parameters of Nature, but that rather they depend on the details of the particular vacuum. This concept is not just idle speculation: it is suggested by superstring theory, which appears to have an enormous landscape of vacua [ $\mathcal{O}(10^{272,000})$ by one recent estimate~\cite{Taylor:2015xtz}], with no known selection principle. Different vacua involve
different gauge groups, hierarchies, parameters,  remnants, and number of large dimensions, and only a subset would be habitable (just right).
Furthermore, eternal inflation~\cite{Linde:2015edk} could have sampled these vacua, creating a multiverse
of spatially separated universes with different physics.\footnote{Variations on this idea could involve different vacua being separated in time, as different branches of a quantum wave function, etc.} Such a string multiverse could provide a reasonable anthropic explanation of some apparent fine-tunings, while other arbitrary features of Nature (e.g., the masses of the third family fermions or the neutrino mixings) could be essentially random.\footnote{It would not necessarily mean that \emph{all} aspects of the SM are arbitrary.}

\subsubsection*{Is string theory and the multiverse verifiable?}
The major problems with this string theory/eternal inflation/multiverse picture are whether it is  testable? Falsifiable? Believable? Desirable? True? Scientific? Many scientists reject it on these grounds. However:
\begin{itemize}
\item Superstring theory is a consistent and finite theory of quantum gravity (at least at the perturbative level). The importance of this should not be forgotten.
\item The multiverse concept is not ad hoc: it may eventually be shown to follow as a necessary consequence of string theory and eternal inflation.
\item Some types of new physics that are allowed in field theory occur rarely or never in superstring vacua.
Examples include large representations, global symmetries,  and tri-fundamental representations. Their observation would go far towards falsification. Recently, similar themes have been expressed in the ``swampland'' conjectures (e.g.,~\cite{{Brennan:2017rbf}})  that most otherwise-consistent effective field theories 
are probably inconsistent with  theories of quantum gravity such as the string landscape. These conjectures are well-motivated by string constuctions.
\item String remnants are common in the landscape~\cite{Halverson:2018xge}, and their observation would provide positive support.
\item Technical progress is possible. For example, it is currently debated whether the landscape includes many or any vacua that are metastable or de Sitter (positive energy), or how many correspond to quintessence, but these questions should essentially be settled. Similarly,  theory and observations should clarify the status of eternal inflation.
\item Completely new ideas may emerge. Who would have anticipated the role of the CMB for cosmology, or of
DNA for paleoanthropology?
\end{itemize}

My view is that the multiverse should be taken very seriously as a possibility.
Its establishment would be the sixth Copernican revolution!\footnote{The Earth is not the center of the Solar System, or of the galaxy. There are other stellar/planetary systems, and other galaxies. Ordinary matter constitutes only a fraction of the total energy density.}

\subsection{How will we make progress?}
There are many ideas for BSM physics, and of course there could be things that no one has thought of.
Fortunately, progress is anticipated on a number of fronts.

\subsubsection*{The energy frontier}
The LHC program to date has been spectacular. The accelerator, detectors, computing, and theoretical support
(QCD calculations, simulations, lattice calculations, \ldots) have all been impressive, and the program has functioned as a model of international cooperation. Physics highlights include verifications of the SM, the Higgs discovery, and numerous constraints on BSM.

\begin{figure}[htbp]
\begin{center}
\includegraphics*[scale=0.45]{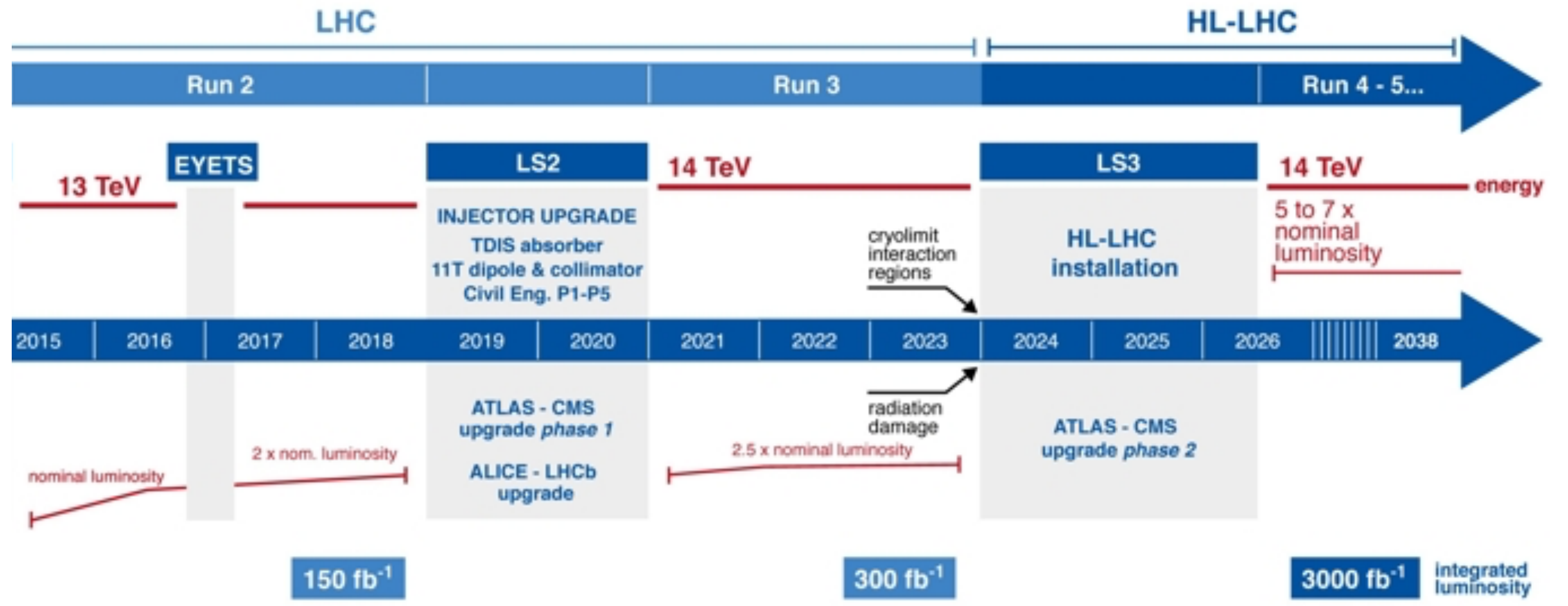}
\includegraphics*[scale=0.42]{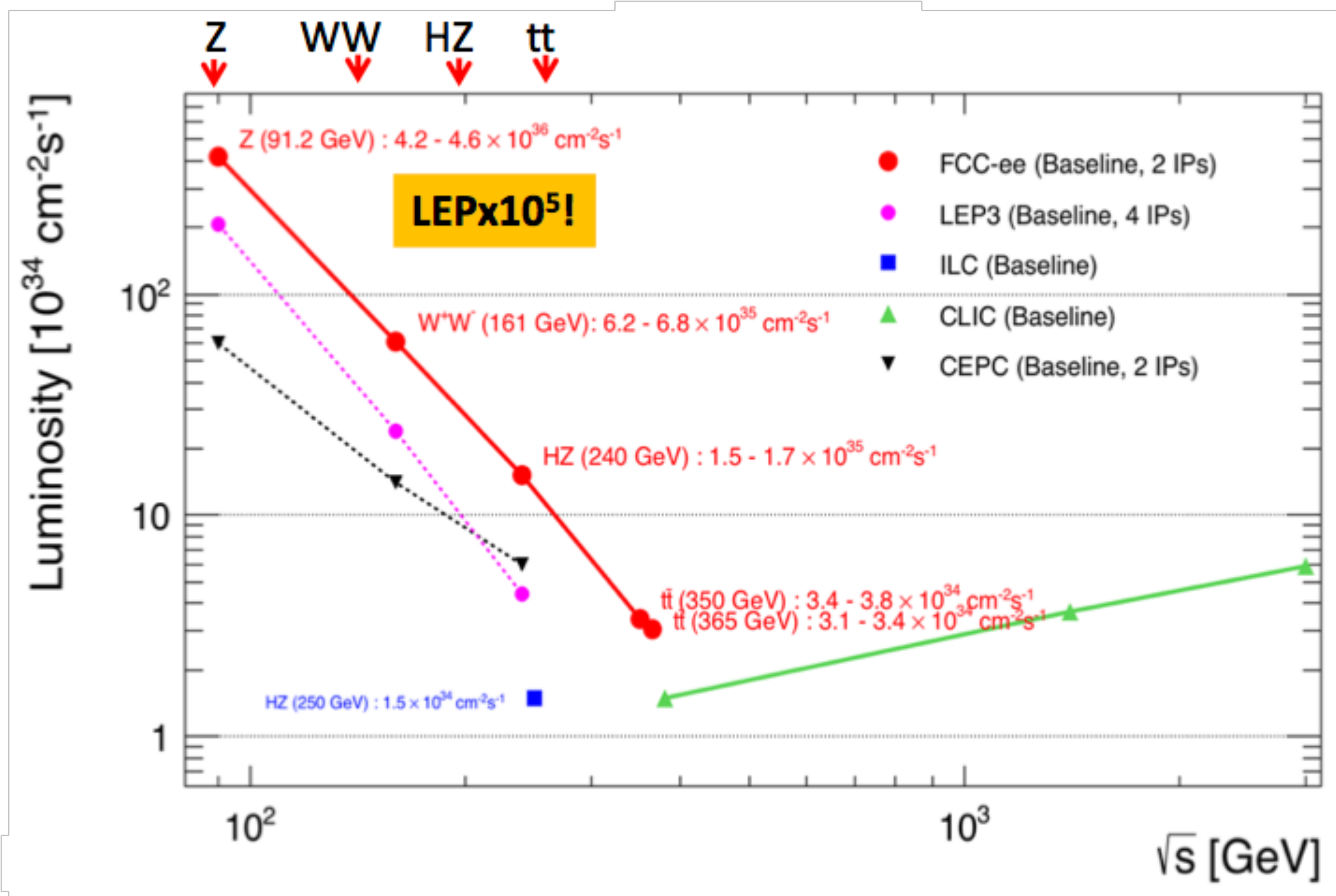}
\caption{Top: LHC luminosity~\cite{Rahatlou}. Bottom: proposed $e^+e^-$ colliders~\cite{Blondel}.}
\label{future}
\end{center}
\end{figure}

The next 20 years or so will witness a large increase in luminosity [Figure~\ref{future}], allowing more precise measurements of the Higgs couplings, searches for BSM, dark matter, and remnants.

On a still longer time scale there are serious proposals for much higher energy [$\mathcal{O}$(100 TeV)]
$pp$ colliders (the FCC-$hh$ at CERN and the SppC in China), which would be able to probe the Higgs, naturalness, dark matter, remnants, and electroweak symmetry breaking and baryogenesis. Other possibilities at the energy frontier include the  proposed HE-LHC (at $\sim$27 TeV), electron-hadron colliders (LH$e$C,  FCC-$eh$), and a muon collider (FCC-$\mu\mu$).  However,  there is no ``no-lose'' theorem for these proposals
analogous to that at the LHC for electroweak symmetry breaking.

There is also expected to be a vigorous continuation of the heavy ion program, with implications for the QCD phases, including the quark-gluon plasma.

Finally, new technologies are being pursued for detectors and accelerators, including superconducting magnet technologies, plasma wave acceleration, etc. New techniques in machine learning for data analysis are being pursued. These could also find application in studying the string landscape.

\subsubsection*{The precision frontier}
There are a number of proposed $e^+e^-$ colliders [Figure~\ref{future}]. 
The International Linear Collider (ILC) in Japan would initially run at 250 GeV, but  could be upgraded to higher energies.
The FCC-$ee$ and CEPC are circular collider proposals that would have higher luminosity but would be limited in energy because of synchrotron radiation. These would be the first stages of the 100 TeV $pp$ collider projects in CERN and China, respectively. The CLIC proposal at CERN would be a very high energy linear collider with a novel drive-beam acceleration technology.  
The $e^+e^-$ colliders could carry out precise studies of the Higgs and search for BSM. They could also repeat the
$Z$-pole program with much higher precision, and determine the $W$ and $t$ masses more accurately.

There are many large and small flavor physics experiments, including the  BELLE II and LHCb programs and smaller 
dedicated experiments on $g_\mu-2$, FCNC, kaon physics,  EDMs, etc.; new large-scale proton decay experiments HyperKamiokande and DUNE~\cite{Yu}; and searches for 
weakly coupled, light, dark, or long-lived particles at colliders, beam dumps, AMO, and dedicated experiments.

\subsubsection*{Neutrino physics}
There are numerous outstanding questions in neutrino physics. These include  the details of the spectrum (leptonic $CP$ violation, mixings, the mass hierarchy) and whether it is due to a symmetry or anarchy; whether the masses are Majorana or Dirac, and their origin; light  sterile neutrinos and their mixing with ordinary neutrinos;  keV-scale sterile neutrinos; electromagnetic properties and nonstandard interaction; decays; Solar physics; supernova physics; and high energy sources.

There is a large world-wide program to address these issues, including long and short baseline accelerator experiments; reactor, source, and spallation experiments; under ice and under water experiments; Solar and atmospheric neutrinos; cosmology; $\beta$ decay and related experiments;   $\beta\beta_{0\nu}$; and (possibly) detection of relic neutrinos.

\subsubsection*{Cosmology and astrophysics}

Topics most directly connected to particle physics include inflation or alternatives; dark matter and energy; constraints on the mass, number, and interactions of ordinary and sterile neutrinos, axions and ALPs, and other exotic particles; primordial black holes, moduli, and defects (e.g., cosmic strings, domain walls, monopoles); phase transitions; and exotic matter at high temperature or density.

We are fortunate to be  living in a golden age of observations in astrophysics and cosmology.
Relevant  programs\footnote{Analyses are  complicated by the discrepancy between local and CMB determinations of the Hubble constant.}  include dark matter searches (direct, indirect, collider; axion and ALP searches; and dark photon searches); studies of the CMB polarization, baryon acoustic oscillations, 21 cm radiation, the Lyman-Alpha forest,
 and the DES and LSST telescopes; gravity wave interferometers and multi-messenger astronomy; high energy cosmic rays, $\bar p$, $e^\pm$, $\gamma$, $\nu$;  and very large telescopes.

\subsubsection*{Realities}
There have been outstanding recent  achievements in particle physics, astrophysics, and cosmology, and the
technical prospects for the future are brilliant. However, we are  challenged by daunting political, economic, and human resource realities. It is imperative that as a field we concentrate not only on our science, but also on
outreach and education (to the public, politicians, industry, and especially young people);
diversity; and spinoff technology. We must also emphasize that high energy physics (and big science generally) is a model for
international cooperation.

\subsection{The quest}

It is my hope that we will eventually develop and establish a New Standard Model of Nature, incorporating physics all the way to the Planck scale, and describing the Universe on the largest scale and earliest time (\emph{or} to go as far as possible towards these goals).

Such a lofty aspiration will require significant progress in experiment, observation, and bottom-up  and top-down theory, as well as
ingenuity, hard work, and luck. It will also need the cooperation of Nature.

This goal is extremely ambitious. We may never achieve it, or it may take a very long time, but we should try.

\section{Thanks}
It is a pleasure to thank everybody involved for a wonderful conference. This includes the co-chairs Jihn E. Kim, Un-ki Yang, and Yeongduk Kim; the program committee Young Joon Kwon, Sung Youl Choi, and Suyong Choi; as well as the International Advisory Committee, the local staff members, all of the sponsors, and especially the participants.

\input{bibfile}

\end{document}